\documentclass[sn-mathphys,Numbered]{sn-jnl}
\usepackage{graphicx,subfigure}%
\usepackage{multirow}%
\usepackage{amsmath,amssymb,amsfonts}%
\usepackage{amsthm}%
\usepackage{mathrsfs}%
\usepackage[title]{appendix}%
\usepackage{xcolor}%
\usepackage{textcomp}%
\usepackage{manyfoot}%
\usepackage{booktabs}%
\usepackage{algorithm}%
\usepackage{algorithmicx}%
\usepackage{algpseudocode}%
\usepackage{listings}%
\usepackage{todonotes}

\usepackage{lineno}
\usepackage{siunitx}

\title[Article Title]{SeqSeg: Learning Local Segments for Automatic Vascular Model Construction}

\author[1]{\fnm{Numi} \sur{Sveinsson Cepero}}\email{numi@berkeley.edu}

\author[1]{\fnm{Shawn} \sur{C. Shadden}}\email{shadden@berkeley.edu}

\affil[1]{\orgdiv{Department of Mechanical Engineering}, \orgname{University of California, Berkeley}, \orgaddress{ \city{Berkeley}, \postcode{94720}, \state{CA}, \country{USA}}}

\begin{document}

\abstract{
Computational modeling of cardiovascular function has become a critical part of diagnosing, treating and understanding cardiovascular disease. Most strategies involve constructing anatomically accurate computer models of cardiovascular structures, which is a multistep, time-consuming process. To improve the model generation process, we herein present SeqSeg (sequential segmentation): a novel deep learning based automatic tracing and segmentation algorithm for constructing image-based vascular models. SeqSeg leverages local U-Net-based inference to sequentially segment vascular structures from medical image volumes. We tested SeqSeg on CT and MR images of aortic and aortofemoral models and compared the predictions to those of benchmark 2D and 3D global nnU-Net models, which have previously shown excellent accuracy for medical image segmentation. We demonstrate that SeqSeg is able to segment more complete vasculature and is able to generalize to vascular structures not annotated in the training data.
}


\maketitle

\section{Introduction}

\begin{figure}[h!]
\begin{center}
\includegraphics[width=\linewidth]{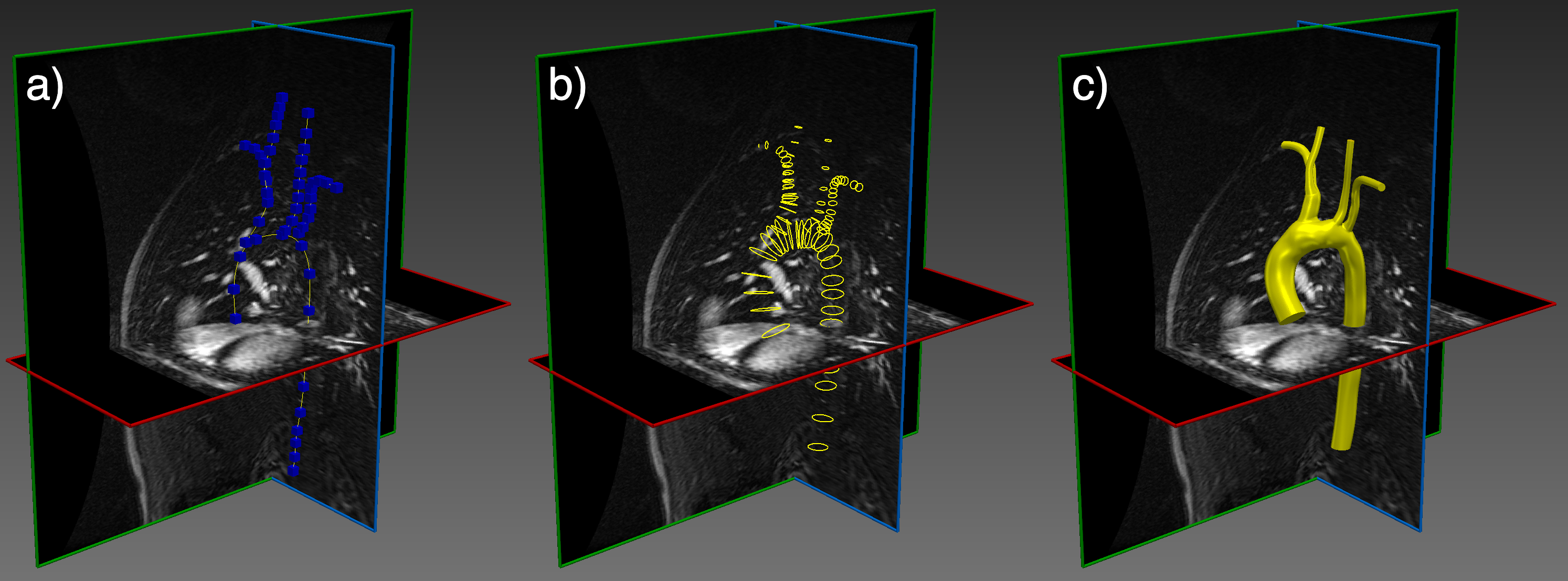}
\end{center}
\caption{A typical vascular model construction workflow involves (a) creating vessel paths by manual selection of point (b) sequential segmentation of the vessel lumen boundary at discrete cross-sections along the paths and (c) lofting these segmentation rings into a unified model. This process is described in more detail in \citep{Updegrove2017SimVascular:Simulation}.
}\label{fig:simvasc}
\end{figure}

Image-based vascular modeling is used for a variety of purposes including diagnosis, personalized treatment planning and fundamental understanding of disease progression \citep{Mukherjee2018TheStudy,Mirramezani2020AFlow,Arzani2014EffectMixing,Sengupta2012Image-basedDisease}. Specialized software has been developed for such modeling, including SimVascular~\citep{Updegrove2017SimVascular:Simulation, Lan2018APackage}, CRIMSON~\citep{Arthurs2021CRIMSON:Simulation} and VMTK~\citep{Izzo2018TheImages}. This modeling paradigm uses medical imaging, such as computed tomography (CT) or magnetic resonance (MR) angiography, to construct a patient-specific anatomical model of vessels of interest. This geometric model is subsequently converted into a 3D computational mesh to support detailed blood flow and/or tissue mechanics simulation and analysis. The construction of an anatomical model from medical image data remains largely a manual process~\citep{Updegrove2017SimVascular:Simulation}. Figure~\ref{fig:simvasc} shows a typical workflow for vascular model construction, starting with the creation of centerlines along the vessels of interest, 2D segmentation of the vessel lumen along the centerlines, and lofting of the 2D segmentations to generate a unified 3D model of the vascular geometry. Alternative segmentation approaches exist, including region-growing or level-set methods~\citep{Lan2018APackage}; however, these methods generally struggle in the segmentation of highly-branched structures such as blood vessels, particularly in the context of limited image resolution, unclear boundaries and image artifacts~\citep{Moccia2018BloodMetrics}. Additionally, when the model is constructed manually, substantial user bias may result.  Ultimately, despite the popularity and maturity of image-based cardiovascular modeling over the past 20 years, the process of deriving a simulation-suitable anatomical model from medical image data has remained a primary bottleneck for large-cohort studies or translational applications where timely results are needed. 

Recently, machine learning has been applied to automate and speed up image segmentation. Note that while medical image segmentation is performed for a variety of healthcare applications, we focus here on the purpose of generating a {\em simulation-suitable} model that can be utilized to generate a computational domain for physics-based simulation. Simulation suitable models have certain criteria that must be met such as, being connected,  sufficiently ``smooth'', and able to be meshed (discretized) with quality elements. Most learning methods focus on pixel classification, which often results in segmentations that are disconnected or have substantial artifacts that complicate, or prevent, generation of a mesh suitable to support numerical simulation. 

Most progress has been made when machine learning has been applied to isolated anatomic vascular regions~\citep{Nazir2020OFF-eNET:Segmentation,Chen2012AutomaticImages} including for cardiac models~\citep{Kong2020AutomatingSimulation,Kong2021AReconstruction}. In the work of Maher, et al.~\citep{Maher2019AcceleratingNetworks,Maher2020NeuralModeling} segmentation of branched vascular domains was achieved by assuming the existence of vessel centerlines (cf. Fig.~\ref{fig:simvasc}a). Under such assumptions, these centerlines are traversed and local 2D cross-sectional segmentations of the lumen boundary are generated using a trained network. This framework essentially automated step (b) shown in Fig.~\ref{fig:simvasc}. However, for many vascular models, the generation of vessel centerlines is the most labor intensive step. Moreover, with this approach, segmentation is only performed at discrete 2D slices along the vessel, which provides incomplete sampling and can be problematic when the cross-section is not connected or the centerline is not sufficiently aligned with the vessel. And more importantly, discrete cross-sectional segmentation performs poorly at vessel bifurcations, which are present in almost all applications. 

\begin{figure}[h!]
\begin{center}
\includegraphics[width=1\linewidth]{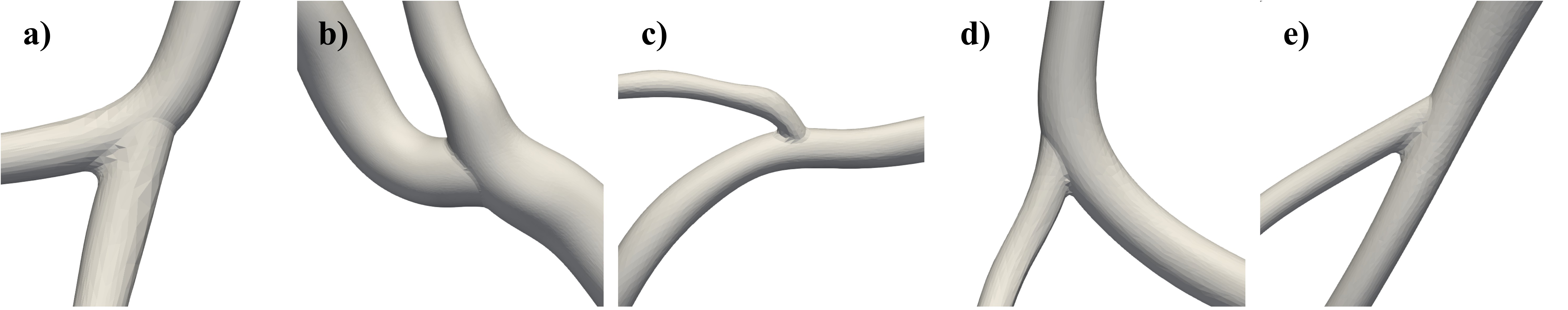}
\end{center}
\caption{When viewed locally, vasculature of different sizes and anatomical regions exhibit substantial geometric similarity. A) the pulmonary artery ($r=1.5mm$), b) the brachiocephalic artery ($r=9mm$), c) the coronary artery ($r=1mm$), d) the cerebral artery ($r=2mm$) and e) the femoral artery ($r=3mm$) are presented}\label{fig:sim}
\end{figure}

Herein, we present a novel method for segmenting branched vascular geometries from medical image data utilizing local deep learning-based segmentation that does not require {\em a priori} centerline information. This approach starts from a seed point and generates a local 3D segmentation of the vessel(s) containing the seed point over a local subvolume. Based on this local segmentation, we determine the orientation of the vessel and any locally connected branches. We then step the subvolume along the determined vessel direction (and new subvolumes along the identified local branch directions) to generate a 3D segmentation of the neighboring segment(s). This approach is motivated by the following considerations: when viewed locally by a subvolume that is centered on a vessel and slightly larger than the vessel diameter, vessels of different sizes and from different regions exhibit substantial geometric similarity (Figure~\ref{fig:sim}), and consequently learning to locally segment a portion of a vessel should be easier than learning to segment an entire vascular network. 
While cropping of medical image volumes has been performed previously, for example, for coronary tracking \citep{Wolterink2019CoronaryClassifier} \citep{Li2021AExtraction}, to the best of our knowledge such approach has not been used to generate 3D segmentation or for segmentation of general vascular geometries.

By processing local subvolumes, we simplify the deep learning task and introduce beneficial inductive bias to the machine learning model, allowing it to generalize to vasculature not present in training data. We test this method on a dataset of publicly-accessible aortic and aortofemoral models, and the results are compared to benchmark global 2D/3D nnU-Net neural network models that have previously shown excellent results for medical image segmentation. The main contribution of this work is a new method capable of:
\begin{itemize}
    \item Tracing vasculature after initialization with a single point and vessel radius estimate.
    \item Segmenting vasculature while ensuring global connectivity to maintain physiologic topology.
    \item Detecting bifurcations, storing them and tracing them sequentially.
    \item Delivering a global surface mesh of segmented vasculature.
    \item Generalizing to segment parts of vasculature not annotated in training data.
\end{itemize}

\section{Method}

\subsection{Algorithm}
\begin{figure*}[h!]
\begin{center}
\centerline{
\includegraphics[width=\linewidth]{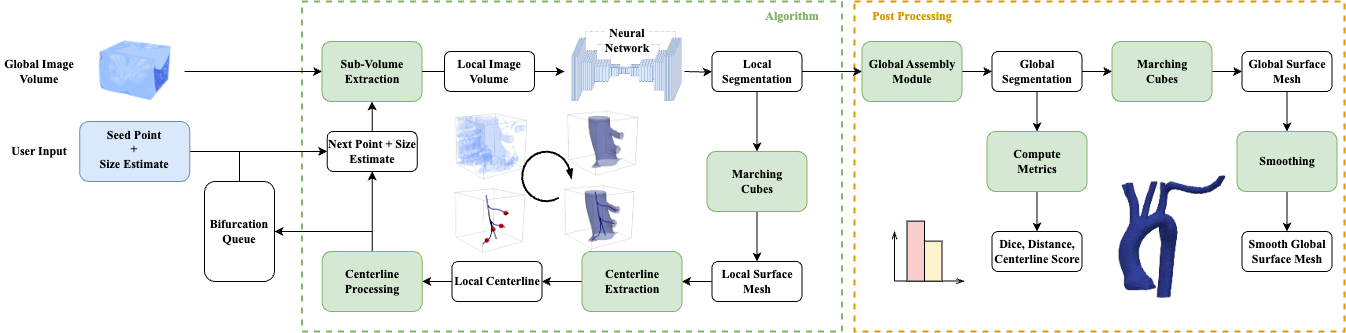}

}
\end{center}
\caption{Overview of the tracing and segmentation algorithm with inputs of the global raw image and seed points for initialization. The algorithm takes steps, stores bifurcations in the queue during tracing, and outputs a global segmentation map for post processing }\label{fig:algo}
\end{figure*}

Figure~\ref{fig:algo} shows a schematic of the algorithm. Breifly, a ``seed point'', (specifying a location and direction) and a rough diameter ``size estimate'' of the vessel containing the seed point are supplied by the user. A local subvolume surrounding the seed point is extracted from the global image volume. The vessel portion contained in the subvolume is segmented using a neural network. The segmentation is postprocessed and converted to a surface mesh, after which a centerline is extracted. The resulting centerline is subsequently used to choose the next subvolume location and size. These steps are explained in further detail below. 

\subsection{Segmentation}

\subsubsection{Dataset, Sampling and Augmentations} \label{sec:sampling}
To train the U-Net and test the algorithm, we utilized a dataset of 41 CT and 44 contrast enhanced MRI aortic and aortofemoral cases, which is commensurate with the amount of data typically provided in segmentation challenges. The breakdown of how many cases are used for training, validation and testing is specified in Table~\ref{tab:dataset}. The VMR datasets used for training are accessible from the open access Vascular Model Repository at \url{https://vascularmodel.com}. For further testing on CT images, we also use a subset of the AVT dataset~\citep{Radl2022AVT:Masks}, specifically the dataset obtained from Dongyang Hospital. Table~\ref{tab:dataset} shows details on the datasets; modalities, purpose, training/test split, anatomies, diseases (if present), sex ratio and age ranges. The datasets contain a 3D image volume and a respective ``ground truth'' vascular segmentation map (converted from segmentation surface models for the VMR data) and corresponding centerlines that served as ground truth labels for training and testing.  

\begin{table}
\centering
\caption{The datasets of patients used for model training and method testing. Abbreviation are as follows: Datasets; VMR: Vascular Model Repository, AVT-D: Aortic Vessel Tree dataset, subset from Dongyang Hospital. Anatomy; AO:Aorta, AF:Aortofemoral. Disease; H:Healthy, AAA:Abdominal Aortic Aneurysm, MA: Marfan Syndrome, CA:Coarctation of Aorta, AOD:Aortoiliac Occlusive Disease, SVD:Single Ventricle Defect. Sex; M:Male, F:Female, U:Unknown. Sex and age information was not available for the AVT dataset.}
\begin{tabular}[t]{lccccccc}
\toprule
Dataset & Modality & Purpose & Train/Test & Anatomy & Disease & Sex & Age(yr)\\
\midrule
VMR & CT & Train/Test & 33/8 & 25 AO, & 23 H, & 23M  & 6 - 80\\
 & & & &16 AF & 15 AAA, & 6F & ave: 58 \\
 & & & & &  3 MA & 12U &\\
 \midrule
 VMR & MR & Train/Test & 37/7  & 38 AO, & 19 H, 14 CA, & 30M & 0.6 - 67 \\
 & & & & 6 AF &  5 SVD, 2 MA & 14F & ave: 17\\
 & & & & &  4 AOD & &\\
\midrule
AVT & CT & Test & 0/18 & 18 AO & 18 H & - & - \\
-D\citep{Radl2022AVT:Masks} & \\
\end{tabular}
\label{tab:dataset}
\end{table}

To generate training data for the local segmentation U-Net, the global 3D medical image volumes in the VMR training datasets were sampled along the centerlines and these subvolumes (i.e. ``Patches'') were stored. Namely, two volumes were extracted at each Patch: 1) the original medical image data and 2) a binary segmentation of the subvolume based on the model representing the ground truth label.

\begin{figure*}[h!]
\begin{center}
\centerline{
\includegraphics[width=\linewidth]{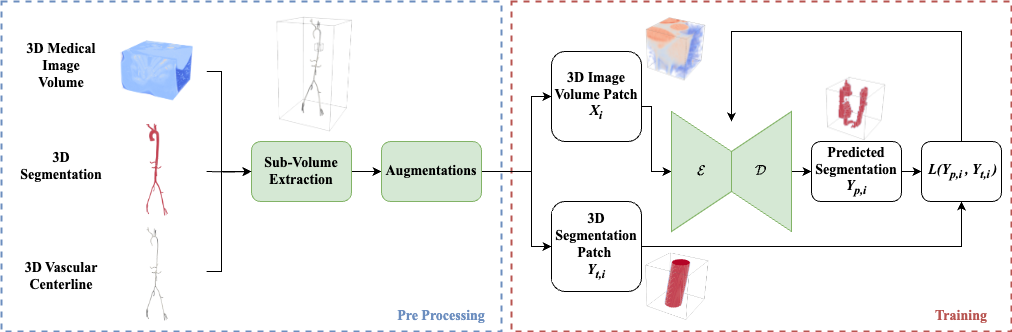}
}
\end{center}
\caption{ Preprocessing involves extracting subvolumes along ground truth centerlines and data augmentation prior to neural network training. Thousands of samples are acquired from only a few dozen models. The neural network consists of an encoder $\mathcal{E}$ followed by a decoder $\mathcal{D}$, which outputs the predicted segmentation map used to compute loss, \textit{L}, during training }\label{fig:Pre}
\end{figure*}

To improve the learning process, we varied the samples in terms of centering and size. Briefly, some samples were centered along the centerline while others were shifted from the centerline, and the subvolume sizes varied from just capturing the lumen of the vessel to including more surrounding tissue. More specifically, each sample volume $s_i$ is a cube dependent on its side length and center, i.e., $s_i(L_i, \textbf{c}_i)$ where $L_i$ is its side length and $\textbf{c}_i$ is the center point of sample $i$. The side length and center are sampled as follows:
\begin{equation}
\begin{aligned}
    & L_i = R_i * \alpha_i \\
    & \textbf{c}_i = \textbf{C}_i + \beta_i * R_i * \textbf{w}_i \\
    & \alpha \sim \mathcal{N}(\mu_r,\,\sigma_r^{2}) \\
    & \beta \sim \mathcal{N}(\mu_s,\,\sigma_s^{2}) \\
\end{aligned}
\end{equation}
where $R_i$ is the local radius of the vessel, $\textbf{C}_i$ is the point on the centerline, $\textbf{w}$ is a unit vector perpendicular to the centerline and $\alpha, \beta$ represent the radius ratios used to enlarge or shift the sample. $\textbf{w}$ was chosen by sampling a random linear combination of orthogonal unit vectors $\textbf{u}, \textbf{v} $ that defined a plane perpendicular to the centerline:
\begin{equation}
\begin{aligned}
    \textbf{w}_i = \frac{a_i*\textbf{u} + b_i* \textbf{v}}{\| a_i*\textbf{u} + b_i* \textbf{v}\|} \mbox{,}\quad a, b \sim U[-1,1]
\end{aligned}
\end{equation}
where $a,b$ are scalars sampled from a uniform distribution between $[-1,1]$.
We used $(\mu_r,\,\sigma_r^{2})=(5,1) $ and $(\mu_s,\,\sigma_s^{2}) = (0, 0.8) $ so that the mean sample was $5$ times the size of the radius and centered on the centerline. This stochasticity was purposefully added to represent the variance that the tracing algorithm encounters during inference and is intended to increase the robustness of the neural network. This process resulted in a training dataset $D$ consisting of $N$ pairs of image subvolumes $X_i$ and corresponding blood vessel segmentations $Y_{t,i}$:
\begin{equation}\label{eq:dataset}
    D = \{(X_1,Y_{t,1}),(X_2,Y_{t,2}), ..., (X_N, Y_{t,N}) \}
\end{equation}
In total, we get $D = 36289$ patches for CT data and $D = 33603$ patches for MR data. For the VMR dataset 15 patient datasets were excluded for final testing, i.e., 8 CT and 7 MR cases were not sampled for training or validation. The generation of subvolume data for network training is shown schematically in Fig.~\ref{fig:Pre}.

Before training, MR image volumes were normalized via z-scoring, where each voxel value, $x$, is subtracted from the image mean $\mu$ and then divided by the image standard deviation $\sigma$. CT volumes were clipped and z scored according to foreground image values where $\mu$ and $\sigma$ are calculated only from voxels labelled as vessel in the ground truth training data and held constant during inference~\cite{Isensee2021NnU-Net:Segmentation}; see Table~\ref{tab:nnunet} for details. The preferred image spacing was chosen as the median spacing across all cases and all image volumes were resampled using a 3rd order spline. Segmentation maps were resampled differently, using linear splines on one-hot encoded maps, and then the argmax of the result.

\subsubsection{Neural Network Architecture and Training}
The convolutional neural network (CNN) U-Net architecture was used for segmentation. The U-Net is tailored to the processing of medical images, going from the original resolution down to a low-resolution, high-dimensional space and then back up to the original resolution. Herein, a 3D version of U-Net was chosen for the SeqSeg segmentation step. Note, we also compare our end results to those of two benchmark U-Net models, i.e. a 2D U-Net and a 3D U-Net, trained on the global image volumes, see Section~\ref{sec:exp} for further details. The U-Net learns features primarily through two different mechanisms. First, by downsampling the original image data the model is forced to retain only important global information when squeezed through a lower-dimensional space. Second, by using skip connections across the neural network the model is able to retain features related to finer details from the higher resolution image in its final prediction. The skip connections are concatenations of blocks of the same resolution. The final output is a pixelwise probability map indicating the likelihood that a pixel corresponds to a target tissue.

Our U-Net was trained for binary classification: to predict whether voxel $y_{i,j,k}$ inside segmentation mask $Y_p \in \mathbf{R}^{W\times H\times D}$ belongs to a blood vessel:

\begin{equation} \label{eq:Y_p}
    \begin{aligned}
        & Y_p = \{y_{i,j,k} \in [ 0,1 ] \mid 0\leq i < W ; 0\leq j < H ; 0\leq k < D\} \\
        & y_{i,j,k} = \left\{ 
  \begin{array}{ c l }
    1 & \quad \textrm{if belongs to vessel} \\
    0 & \quad \textrm{otherwise}
  \end{array}
    \right.
    \end{aligned}
\end{equation}
where $i,j,k$ refers to the index of a voxel in an image of width $W$, height $H$ and depth $D$. In this section, lower case notation refers to individual nodes or voxels, e.g. $y$, whereas capital notation, e.g. $Y$, refers to a set of nodes or voxels such as composing an image, segmentation mask or output from neural network layers.

In mathematical terms, the neural network is a parameterized function $f_{\theta}$ that transforms a raw image input $X \in \mathbf{R}^{W\times H\times D}$ into a blood vessel segmentation map:
\begin{equation}
    Y_p = f(X \mid \theta )
\end{equation}
where $\theta$ are the parameters of the neural network, which are optimized using training data. The final output, $Y_p$, ranges between $[0,1]$ and can thus be interpreted as a probability map of whether each voxel belongs to a blood vessel. This enables the volume to be binarized by thresholding to a particular probability value. 

We utilized the nnU-Net framework for hyperparameter specification and training \citep{Isensee2021NnU-Net:Segmentation}. The framework automatically determines parameters such as image resampling spacing, patch size and batch size based on training data and GPU memory size. The underlying neural network architecture used is the U-Net, with additional constraints on specific parameters. Table \ref{tab:nnunet} lists the specifications of our implemented U-Net model architectures and training parameters. Since the SeqSeg model is trained on smaller volumes compared to the benchmarks, its required batch size can be larger, see Table \ref{tab:nnunet}. The nnU-Net framework utilizes stochastic gradient descent with Nesterov momentum with an initial learning rate of $0.01$ accompanied by a learning rate scheduler of $(1 - epoch/epoch_{max})^{0.9}$, where $epoch_{max} = 1000$ was chosen; see \citep{Isensee2021NnU-Net:Segmentation} for further details. Training was performed using an NVIDIA Geforce RTX 2080ti GPU (11 GB GPU memory) on the Savio High Performance Computing cluster at the University of California, Berkeley. 

\begin{table}[h]
\caption{The U-Net architecture and training specifications, for both the SeqSeg models and global benchmark models }
\centering
\begin{tabular}{c|cccccc}
    {Parameter} & {SegSeg CT} & {SeqSeg MR} & {3D CT} & {3D MR} & {2D CT} & {2D MR}\\ 
    \midrule
    Intensity  & 0.5/99.5\% & all image  & 0.5/99.5\% & all image & 0.5/99.5\% & all image \\
    Normalization & clip +  & z-score & clip + & z-score & clip + & z-score\\
     & foreground &  &foreground &  &foreground& \\
     & z-score & & z-score & & z-score &\\
    \midrule
    Image  & 0.200,   & 0.0859,   & 0.0800, & 0.0859,   & 0.0488,  & 0.0586, \\
    Target & 0.0547, & 0.0625, & 0.0488, & 0.0586, & 0.0488 & 0.0586\\
    Spacing & 0.0547 & 0.0625 & 0.0488 & 0.0586 & &\\  
    \midrule
    Patch Size   & [20,80,80] & [40,48,48] & [96,160,160] & [56, 256, 160] & [512, 512] & [512,384]\\ 
    \midrule
    Batch Size & 33 & 57 & 2 & 2 & 12 & 16 \\
    \midrule
    Max Nr.  & 320 & 320 & 320 & 320 & 512 & 512 \\ 
    Features  &  &  &  &  &  & \\ 
    \midrule
    Nr. Stages & 5 & 4 & 6 & 6 & 8 & 7 \\
    Encoder  &  &  &  &  &  & \\ 
    \midrule
    Nr. Stages  & 4 & 3 & 5 & 5 & 7 & 6 \\
    Decoder  &  &  &  &  &  & \\ 
    \midrule
    Nr. Layers  & 2  & 2  & 2  & 2 & 2 & 2 \\ 
    per Stage  &  &  &  &  &  & \\ 
    \midrule
    Nr. Pooling & [2,4,4] & [3,3,3] & [3,5,5] & [4,5,5] & [7,7] & [6,6] \\
    Ops. per Axis  &  &  &  &  &  & \\ 
    \midrule
    Conv. Kernel  & [3,3,3] & [3,3,3] & [3,3,3] & [3,3,3] & [3,3] & [3,3] \\
    Size  &  &  &  &  &  & \\ 
    \hline
\end{tabular}
\label{tab:nnunet}
\end{table}


\subsubsection{Loss Function}
The loss function was a combination of Dice score ($
\mathcal{D}$
) and binary cross-entropy ($\mathcal{CE}$):
\begin{equation}
    \mathcal{D}(Y_{p},Y_{t}) =\frac{2 \cdot\|Y_{p} \cap Y_{t}\|}{\|Y_{p}\|+\|Y_{t}\|}
\end{equation}
\begin{equation}
    \mathcal{CE}(Y_{p},Y_{t})=\frac{1}{n} \sum_{y \in Y} \left(y_{t} \cdot \log y_p+\left(1-y_t\right) \cdot \log \left(1-y_p\right)\right)
\end{equation}
where $Y_{p}$ and $Y_{t}$ are respective prediction and ground truth segmentation masks, respectively, and $n$ is the total number of voxels. $Y_t$ is defined similarly to $Y_p$ in Eq.~\eqref{eq:Y_p}. Binary cross entropy is a common loss function for binary classification and we added Dice loss to regulate it for medical image segmentation. Namely, the Dice score helps counter the class imbalance that pixelwise classification problems face in medical image segmentation. This is critical when working with 3D images where the number of voxels belonging to a blood vessel is a small percentage of the total voxels in the volume. It follows that our loss function is defined as
\begin{equation}
    \mathcal{L} = \sum _i^{N_b} ( 1 - \mathcal{D}(Y_{p,i},Y_{t,i}) - \mathcal{CE} (Y_{p,i},Y_{t,i}))
\end{equation}
for a batch size $N_b$, where each batch is a subset of the total dataset $N_b < N$ described in Eq. \eqref{eq:dataset}. The data is batched to fit into GPU memory as described in Table \ref{tab:nnunet}. Each image in the batch is processed in parallel on a GPU and the loss is accumulated before taking a gradient step to update the model parameters.

\subsection{Surface, Centerline Calculations and Step Taking}
As mentioned above, the output of the U-Net is a binarized image subvolume. The marching cubes algorithm~\citep{Lorensen1987MARCHINGALGORITHM} can be applied to this binarized image subvolume to generate a local surface mesh of the vessel segment. The resulting surface was cut using the image subvolume boundary planes, which results in truncation boundaries for the vessel(s), i.e., ``inlets'' or ``outlets''. One of these truncation boundaries is identified as the source (inlet) and others are identified as targets (outlets). This process was performed automatically using information from previous steps and from the direction of tracing. To do this, the centers of the truncation boundaries are calculated. The truncation boundary center closest to the previous stepping point is chosen as source and the rest as targets.

The surface mesh, with respective outlet labels, is used to automatically generate centerline(s) and radius estimates of the local vessel segment using a levelset based centerline extraction method. The method calculates centerline(s) as the path(s) that follow a wave propagation starting from a seed point~\citep{SabryHassouna2005RobustSets}. The wave propagation is modeled by equation:
\begin{equation}\label{eq:levelset}
    |\nabla T(x)|F(x) = 1
\end{equation}
where $T(x)$, the time it takes for wave to reach point $x$, is solved using a set ``speed'' function $F(x)$. $F(x)$ is set to have values proportional to distance from vessel boundary, leading to higher value towards the center and lower closer to vessel walls. When Eq.~\ref{eq:levelset} is solved with $T(x_0)=0$ at source point $x_0$, we obtain a solution with wave propagation faster in the center of vessels. Then, using that solution, we perform gradient descent starting from target point(s), where $T(x)$ is high, until we reach the source point, where $T(x)$ is low, and have therefore defined a centerline path(s). Since the ``speed'' function had higher values towards the center then so do the values of $\nabla T$ which forces the gradient descent towards the center of the vessel while tracing back, see \citep{SabryHassouna2005RobustSets} for details. Furthermore, we estimate the radius of the vessel at each point along the centerline by its distance to the surface.

The centerline extraction depends on well-defined outlet centers fed as seed points. Our method defines these outlet centers automatically, as described above. In the case of a bifurcation, a single outlet was labeled as the source based on the previous step and the direction of tracing. The point(s) to move to along the computed centerline(s) is chosen at $80\%$ along each branch, see stepping point choice in Fig.~\ref{fig:tracing}. We found that this allowed for smaller step size, ensuring more accurate bifurcation detection. 

\subsection{Choice of Subvolume Size and Chances}

\begin{figure*}[h]
\begin{center}
\centerline{
\includegraphics[width=\linewidth]{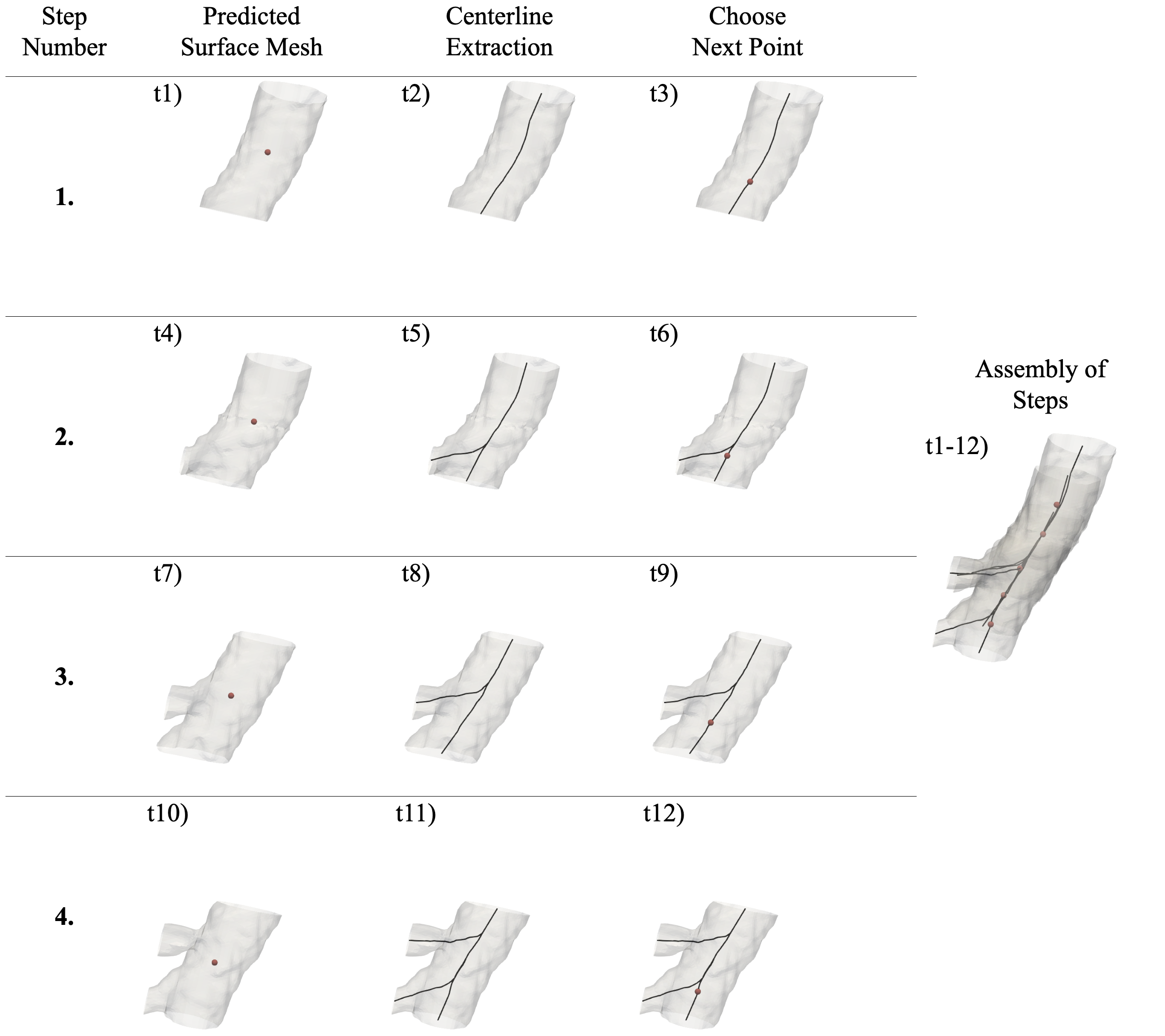}
}
\end{center}
\caption{Automatic tracing using local surface mesh predictions for 3 steps, involving 12 calculation time steps. Centerlines are extracted and the next points are chosen to move to. These steps are subsequently assembled together to form the global vasculature model}\label{fig:tracing}
\end{figure*}

The local centerlines calculated as described above contain 1D (lines) meshes in 3D space connecting all outlets with radius information along them, see Figure~\ref{fig:tracing}. This radius estimate is subsequently used to determine the size of the next subvolume to extract. The length of the next subvolume is chosen as five times the radius estimated, consistent with the size of the training samples as described in Section~\ref{sec:sampling}. Furthermore, to prevent underestimation of subvolume size, we let radius estimate carry on from one step to the next. The subvolume sidelength $L$ is calculatated by an average of the current radius estimate, $r_i$, and the estimate from the previous step, $r_{i-1}$:
\begin{equation}
    L = 5 * (r_i + r_{i-1})/2
\end{equation}
Additionally, we use the segmentation prediction itself as an indicator of subvolume size to vessel size ratio. If a high percentage of voxels within an image subvolume is predicted as belonging to vessel, that may indicate a small subvolume to vessel ratio, i.e., that the vessel occupies a large part of the volume. Utilizing this, we defined a cutoff percentage, $\gamma^*$, for which if the percentage exceeds it then we enlarge the subvolume size until it drops below, see Algorithm~\ref{alg:algorithm_enlarge}, where $X$ is the subvolume, a function of sidelength $L$. 

\begin{algorithm}
\caption{An algorithm to enlarge the subvolume based on the percentage of voxels predicted as vessel, $\gamma$. }\label{alg:algorithm_enlarge}
\begin{algorithmic}

\State $R \gets (r_i + r_{i-1})/2$
\State $L_0 \gets 5 * R$
\State $Y_p \gets f(X(L_0))$
\State $\gamma \gets \frac{1}{W*H*D}\sum_{y \in Y_p} y$
\While{ $\gamma \geq \gamma^*$}
\State $L \gets 5 * R * 1.1$ \Comment{Increase sidelength by 10\%}
\State $Y_p \gets f(X(L))$
\State $\gamma \gets \frac{1}{W*H*D}\sum_{y \in Y_p} y$
\If {$\frac{L}{L_0} > 1.3$}
\State $\textbf{break}$ \Comment{Maximum increase of 30\%}
\EndIf
\EndWhile
\end{algorithmic}
\end{algorithm}

When SeqSeg encounters local subvolumes with image artifacts or unclear vessel boundaries, the neural network model sometimes produces inaccurate segmentations. However, in many cases, these inaccuracies are bound to those specific locations in the image volume whereas the following downstream vasculature may be clearer and easier to segment. To handle these situations, we implemented a ``chances'' feature to SeqSeg. When SeqSeg fails to detect $\geq 2$ outlets or fails to successfully compute a centerline, we give the step another chance and move further in the same direction and try again. Given a point $\textbf{p}_i$ with a corresponding vessel tangent $\textbf{t}_i$ and radius $R_i$, the next ``chance'' location $\textbf{p}_{i+1}$ is calculated as:

\begin{equation}
\textbf{p}_{i+1} = \textbf{p}_i + R * \textbf{t}_i
\end{equation}

We set a maximum number of chances to three. This allows SeqSeg to better move past difficult regions of the image.

\subsection{Bifurcations and Retracing Prevention}

\begin{figure*}[h]
\begin{center}
\centerline{
\includegraphics[width=\linewidth]{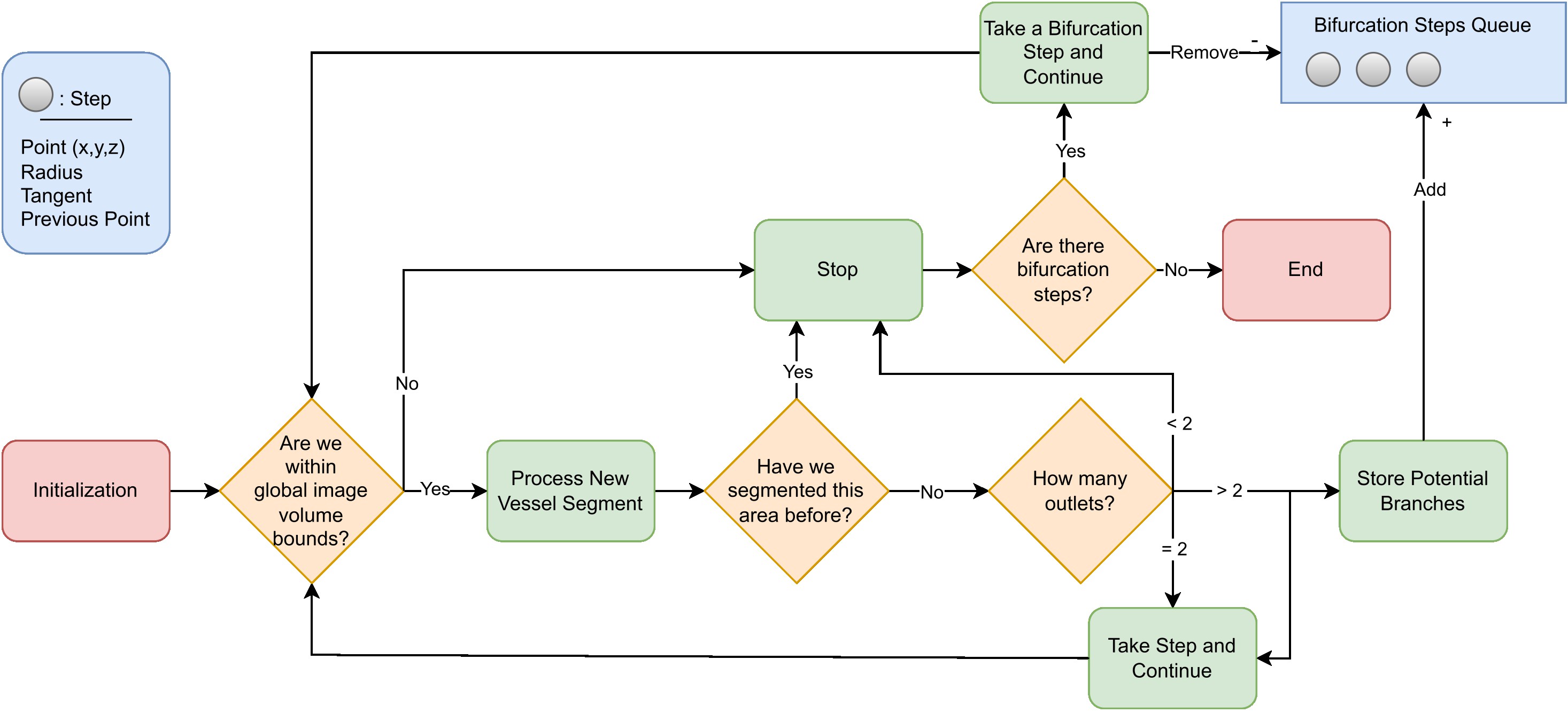}
}
\end{center}
\caption{ How the algorithm takes steps and handles bifurcations, starting from an initial seed point. The bifurcation points are stored in queue for subsequent tracing}\label{fig:logic}
\end{figure*}

Bifurcations are detected by counting the branches of the centerline successfully computed. When bifurcations are detected, they are stored and returned to once other branches have been traced. Namely, the largest radius outlet was chosen for continued tracing while the others were saved as bifurcation points and were revisited once the current vessel had been fully traced. These bifurcation points were periodically sorted by radius to ensure prioritization of the largest vessels first, similar to how a human would interrogate the vasculature; see Figure~\ref{fig:logic}.

Since the method detects outlets locally, it can sometimes detect the same bifurcations multiple times. This occurs especially if a small step size is used to advance the subvolume. To save computational time, we implemented a retracing prevention technique that periodically checks the global segmentation to determine whether the algorithm has segmented the current region before. We added buffers to the global assembly module to ensure that these checks only applied to segmentations involving past branches and not the current one.


\subsection{Initialization}
As mentioned above, the SegSeg method requires a seed point accompanied by a size estimate and direction for initialization. For evaluation purposes, this seed point is chosen at the `start' of each vascular model, in the largest artery closest to the heart, similar to how a user would define it.

\subsection{Stop Criteria}
Since SeqSeg is an automated tracing method, stop criteria are required to terminate step-taking. The current framework has no explicit stop criteria; however, indirectly, it stops when either of two scenarios occur:
\begin{enumerate}
    \item The method reaches the global image boundary, thereby requesting image data that does not exist.
    \item The requested subvolume is of low resolution (e.g., resulting from tracing a small vessel) or is of too low quality (e.g., blurry or has image artifacts) that results in segmentation failure, or a centerline extraction failure from the resulting low-quality segmentation.
\end{enumerate}
In addition to the ``forced'' stop criteria mentioned above, we have implemented \textit{optional} criteria as well, that can be toggled if premature stopping is desired. The optional stop criteria are as follows:
\begin{enumerate}
    \item Define $N_{max}$, maximum number of steps taken and stop once number of steps taken has exceeded $N_{max}$. We set $N_{max} = 500$.
    \item Define $R_{min}$, minimum allowed radius, and stop tracing down a branch once radius estimate is under $R_{min}$. We set $R_{min} = 0.5 \si{mm}$.
    \item Define ${NB}_{max}$, maximum number of branches to be traced down, and stop once number of branches exceed ${NB}_{max}$. ${NB}_{max}$ can, for example, be set as $15$.
\end{enumerate}

\subsection{Global Assembly}
Sampling subvolumes may overlap a given voxel multiple times depending on the step size used to propagate the subvolume along the identified vessel (cf.\ superposition of subvolumes on the right side of Figure~\ref{fig:tracing}), or because of a new subvolume introduced to trace a bifurcation. The end result is that several predictions may exist for a given voxel. Thus, all local segmentations are gathered globally by calculating a weighted mean prediction for each voxel. During development, we noticed that segmentations tend to be less accurate close to the subvolume boundary, so we added Gaussian weighting giving more weight to voxel predictions  closer to subvolume centers. This is also consistent with the benchmark nnU-Net method~\citep{Isensee2021NnU-Net:Segmentation}. For each subvolume prediction of sidelength $L$ and center point $c$, we define a weight map $W$, where each voxel with indices ${i,j,k}$ and point location $p_{i,j,k}$ gets a weight $w_{i,j,k}$ according to:
\begin{equation}
    w_{i,j,k} = \mathrm{e}^{-\frac{\parallel p_{i,j,k} - c\parallel_2}{2\sigma^2}},
    \sigma = \frac{1}{4}L
\end{equation}
Therefore, voxels close to $c$ get a weight close to $1$ while voxels close to the borders of the subvolume get a weight value of approximately two standard deviations from the mean, ~$0.02$.

Given a set $\mathcal{S}$, of size $N_{\mathcal{S}}$, of segmentations in which local volumes included voxel $y_{i,j,k}$, the final global segmentation value was given by
\begin{equation}
    y_{i,j,k} = \frac{1}{N_{\mathcal{S}}\sum_{s \in S}w_{i,j,k}^s}\sum_{s \in \mathcal{S}} w_{i,j,k}^s\cdot s_{i,j,k}
\end{equation}
where $i,j,k$ refer to global voxel indices and $w_{i,j,k}^s$ is the weight value for that voxel associated with segmentation $s$. This was performed prior to thresholding so that the resulting global segmentation retained voxel values ranging between $[0,1]$ depending on the confidence. Finally, the segmentation was upsampled, and thresholded using a value of $t=0.5$:
\begin{equation}
    y_{i,j,k} = \left\{ 
  \begin{array}{ c l }
    1 & \quad \mbox{if}\quad y_{i,j,k} \geq t \\
    0 & \quad \mbox{if}\quad y_{i,j,k} < t
  \end{array}
    \right.
\end{equation}
where the largest connected body is retained, converted to a surface mesh using marching cubes and smoothed to remove voxel artifacts. For mesh smoothing, we use a windowed sinc function interpolation kernel to move mesh vertices\citep{Taubin1996OptimalDesign}. More specifically, we perform ten iterations with a passband value of $0.01$.


\subsection{Experiments, Metrics and Statistical Analysis}\label{sec:exp}
We compared our results to those of two benchmark nnU-Net models, i.e. a 2D nnU-Net and a 3D nnU-Net, trained on the global image volumes. The 3D nnU-Net performs 3D convolutions whereas the 2D version performs 2D convolutions, and outputs 2D predictions, solely on the image z-plane, the patient's axial plane. A 3D segmentation map output from the 2D nnU-Net is assembled by a z-stack of 2D segmentations. The nnU-Net is arguably the most state-of-the-art method for medical image segmentation and thus is chosen as a benchmark for comparison. 

The metrics for comparing SeqSeg with the global nnU-Net benchmarks were as follows:
\begin{equation}
    \mathcal{D}(X,Y) = =\frac{2 \cdot\|X \cap Y\|}{\|X\|+\|Y\|}
\end{equation}
\begin{equation}
    \begin{split}
    \mathcal{H}(X, Y)=\max \{d(X, Y), d(Y,X)\}, \\ 
    \text{where  } \ d(X,Y) = \text{sup}_{x \in X} \text{inf}_{y \in Y} d(x,y) 
    \end{split}
\end{equation}
\begin{equation}
    \mathcal{CO}(Y,C_t) = \frac{\int_{C_t} Y dx}{\int_{C_t} dx}
\end{equation}
where $\mathcal{D}$ is the Dice score, $\mathcal{H}$ is the Hausdorff distance, and $\mathcal{CO}$ is the centerline overlap with $C_t$ being the ground truth centerline and $X$,$Y$ being segmentation maps. Here $X$ represents the ground truth segmentation map and $Y$ the predicted segmentation maps, either output from SeqSeg or the benchmark nnU-Net models. The Dice score measures the overlap between two segmentations and ranges between $[0,1]$. The Dice score is common for medical image segmentation because of its ability to penalize imbalanced datasets accurately. The Hausdorff distance measures the maximum distance between two surfaces and has a minimum of 0 for identical surfaces. The centerline overlap is a score ranging between $[0,1]$ and represents the percentage of the ground truth centerline captured in the predicted segmentation. Both the $\mathcal{H}$ and $\mathcal{CO}$ metrics give additional insight into a method's ability to capture bifurcations and specifically small blood vessels compared to the Dice score. The Dice score compares segmentations on a volumetric basis by comparing voxels, but since most voxels belong to larger blood vessels, it results in indirect bias. 

Not all blood vessels were annotated in the test datasets, or they were present but truncated. We thus masked the outputs from all methods with the volume surrounding the ground truth annotated vessels. We define this mask volume by labelling all pixels within a six radius distance from the ground truth centerline.

For statistical analysis, we perform the Wilcoxon signed-rank test between the resulting metrics scores of SeqSeg and the benchmark. This is a non-parametric test similar to the paired t-test. But since the paired t-test has limitations when comparing machine learning model performances, we opted for the Wilcoxon test~\citep{Rainio2024EvaluationLearning}. Specifically, we test the null hypothesis that the median of differences between the two sets of sample results (metric scores) is zero. A p-value less than $0.05$ was considered to reject the null hypothesis, and therefore indicate a statistically significant difference between the two sets. We both perform Wilcoxon tests between SeqSeg and 2D nnU-Net predictions and between SeqSeg and 3D nnU-Net predictions.

\section{Results}
We tested SeqSeg on 15 held-out VMR image volumes (test set), 8 of which were CT volumes and 7 of which were MR volumes, as well as additional 18 AVT CT image volumes. Namely, SeqSeg and the 2D and 3D nnU-Net benchmark methods were used to segment the vasculature from these image volumes, and those resulting segmentations were compared to manually-generated ``ground-truth'' segmentations from the open data Vascular Model Repository and the AVT dataset\citep{Radl2022AVT:Masks}. 

A quantitative evaluation of the Dice score ($\mathcal{D}$), Hausdorff distance ($\mathcal{H}$) and centerline overlap ($\mathcal{CO}$) for segmentations generated from the VMR test set using SeqSeg and the nnU-Net benchmark methods is presented in Table~\ref{tab:results_nokeep}. SeqSeg, on average, obtained higher Dice scores than the nnU-Net benchmarks in 11 test cases, lower Hausdorff distance scores in 11 cases, and higher centerline overlap scores in 12 cases, all out of a total of 15 test cases. Specifically, SeqSeg on average obtained higher scores than the 2D and 3D nnU-Net benchmarks in terms of Dice score by 0.017 and 0.029, respectively for CT, and 0.036 and 0.029, respectively for MR. For the Hausdorff distance, SeqSeg on average obtained lower distance than the 2D and 3D nnU-Net benchmarks by 0.59 and 0.966 pixels, respectively for CT, and 0.565 and 0.872, respectively for MR data. For centerline overlap, SeqSeg obtained higher scores on average than the 2D and 3D nnU-Net benchmarks by capturing 3\% and 10.9\% more vessel segments, respectively for CT, and 9.4\% and 10.4\% more segments, respectively for MR. Improvements to metric averages that were statistically significant ($p<0.05$) are indicated by an asterisk $^*$ in Table~\ref{tab:results_nokeep}. Beyond mean improvements, SeqSeg appeared far more robust. This can be observed from the the box plots in Figure~\ref{fig:results_vmr_all}, which demonstrates greater consistency in the performance of SeqSeg for all quantitative metrics.

Since the objective of segmentation is a unified, high-quality geometric model, distilling the comparison down to any single metric is overly simplistic (if not deceptive). For broader perspective, Figure~\ref{fig:bestworst_all} provides a qualitative comparison, showing the CT and MR segmentations for which the 2D nnU-Net benchmark yielded the best, median and worst Dice scores, and includes comparison of these to the segmentations generated from SeqSeg. (Note, since the 2D nnU-Net was superior to the 3D nnU-Net, the 2D nnU-Net was considered the de facto benchmark.) This figure demonstrates that SeqSeg generally captures more of the vasculature, and particularly the connections to smaller branch arteries. The Appendix provides a visual comparison of the segmentations generated by SeqSeg and the nnU-Net benchmarks for all CT, and MR, VMR test cases in Figures~\ref{fig:qual_vmr_ct}, and \ref{fig:qual_vmr_mr}, respectively.

\begin{figure}[h]
    \centering
    \subfigure[]{\includegraphics[width=0.48\textwidth]{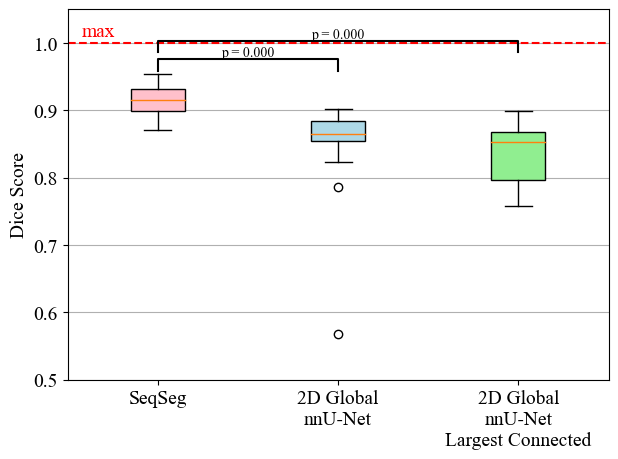}} 
    \subfigure[]{\includegraphics[width=0.48\textwidth]{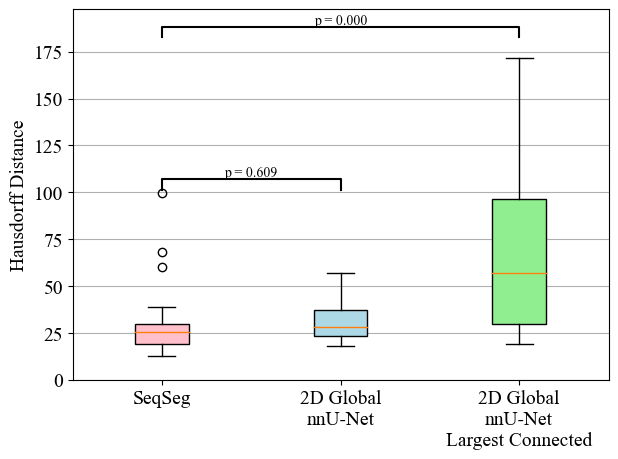}} 
    \subfigure[]{\includegraphics[width=0.48\textwidth]{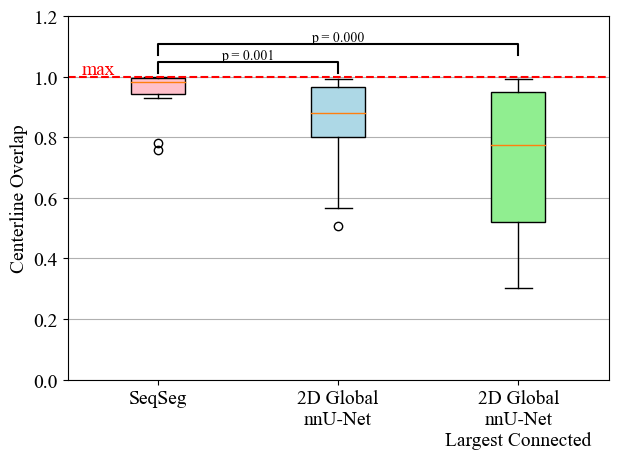}}
    \caption{ Quantitative metric scores for the AVT test dataset for SeqSeg and the 2D nnU-Net benchmark, both raw and after largest connected component filtering. (a) Dice (b) Hausdorff distance (in pixels) (c) Centerline Overlap. See Sec.~\ref{sec:exp} for definition of p-values.
    }
    \label{fig:results_avt}
\end{figure}

For the AVT CT test set, the quantitative metric scores can be seen in Table~\ref{tab:results_avt} between SeqSeg and the 2D nnU-Net benchmark. On average, SeqSeg obtained higher Dice scores by 0.065 and centerline overlap by 10.8\%. On average, the benchmark Hausdorff distance scores were lower by 0.401 pixels. In terms of statistical significance, both Dice and centerline overlap differences were found statistically significant whereas the difference in Hausdorff distance was not. Boxplots of the metric scores are shown in Figure~\ref{fig:results_avt}, again showing a smaller spread (better robustness) for SeqSeg compared to the benchmark, particularly for Dice score and centerline overlap. For qualitative comparison, all resulting meshes are shown in Figure~\ref{fig:qual_avt}. As observed, SeqSeg captures more, and smaller, branches as compared to the benchmark, even branches not included in the ground truth (cf. cases 2, 4, 5, 6, 8, 9, 11, 15, 16, 17 and 18).

Inference time also differed between SeqSeg and the benchmarks. If both are run on the same CPU, measured SeqSeg inference time ranged 20-80min, depending on the number of branches, whereas the nnU-Net benchmarks ranged 2-3hr.

\subsection{Comparison with nnU-Net's largest connected region}

The ultimate goal in image-based modeling is to use a segmentation as the computational domain for numerical simulation.  Simulations require domains to be unified and well defined. Since the nnU-Net segmentations are often disjoint, filtering and keeping only the largest connected body would be necessary to use the segmentation for simulation purposes. Thus, for a more practical comparison, in this subsection we compare SeqSeg segmentations with nnU-Net segmentations that have been filtered to retain the largest connected region.  

We present the quantitative metric values for the largest connected region results in Table~\ref{tab:results_keep}. From this table we can observe that SeqSeg on average obtained higher metric scores than the 2D/3D nnU-Net benchmark, respectively, as follows: the Dice coefficient improved by 0.062/0.032 for CT and 0.064/0.029 for MR; the Hausdorff distance improved by 1.812/2.002 for CT and 2.153/0.839 pixels for MR; and the global centerline overlap increased by 10.2/16.8\% for CT and 18.7/13.8\% for MR. Improvements to metric averages that were statistically significant ($p<0.05$) are indicated by an asterisk $^*$ in Table~\ref{tab:results_keep}. Differences in centerline overlap scores were found statistically significant between SeqSeg and both benchmark methods. The box plots of these metrics for all cases are shown in Figure~\ref{fig:results_vmr_all} and again reveal far less spread in the metrics for SeqSeg compared to both benchmark models, indicating greater robustness in segmentation results for SeqSeg.

\begin{figure*}[h]
\begin{center}
\centerline{
\includegraphics[width=\linewidth]{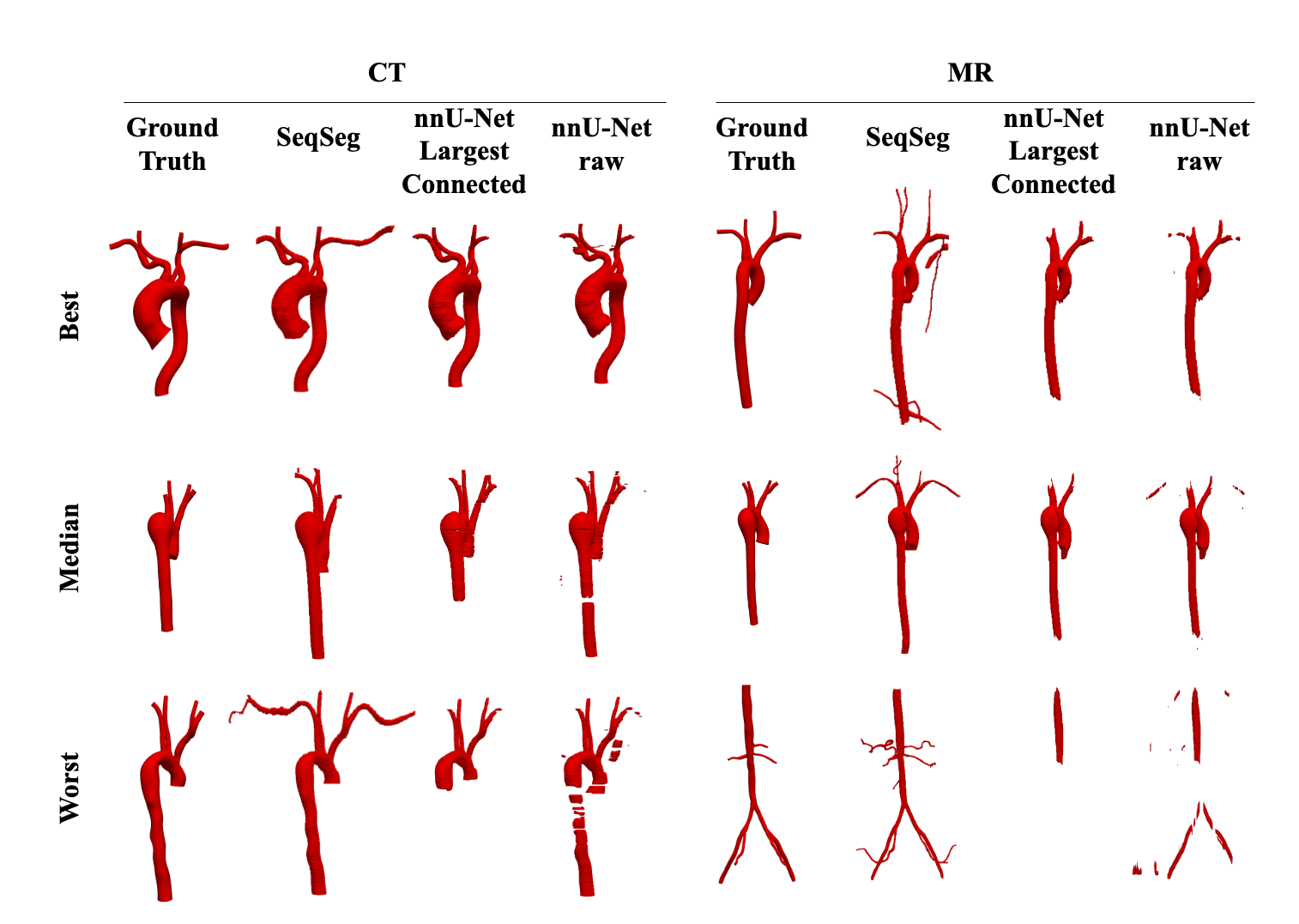}
}
\end{center}
\caption{Qualitative comparison of the resulting meshes on the VRM test dataset, comparing the best, median and worst cases of the nnU-Net benchmark to those of SeqSeg. From Table~\ref{tab:results_nokeep}
these are cases 7, 4 and 5 for CT and 4, 5, and 1 for MR data, respectively}\label{fig:bestworst_all}
\end{figure*}

Figure~\ref{fig:bestworst_all} displays segmentation results for the best, median and worst case results for the 2D nnU-Net benchmark largest connected region, and compares to the segmentation predicted by SeqSeg. (Again, we excluded the 3D nnU-Net in this comparison as it generally performed worse than its 2D counterpart.) For all cases shown, SeqSeg generally captures a greater number of branches and greater extent of the vessels, even when compared to nnU-Net's best results. This is further demonstrated in Figures~\ref{fig:qual_vmr_ct} and \ref{fig:qual_vmr_mr} in the Appendix for all test cases.

For the AVT CT test data, Table~\ref{tab:results_avt} and Figure~\ref{fig:qual_avt} show results for the benchmarks after largest connected component filtering, quantitatively and qualitatively respectfully. We obtain statistically significant difference between SeqSeg and the benchmark for all metrics; Dice, Hausdorff distance and centerline overlap. As shown in Fig.~\ref{fig:qual_avt}, SeqSeg produces better unified vascular trees in more instances than the benchmark.

\begin{table*}[t]
\caption{Quantitative comparisons for the VRM test dataset between the two benchmark U-Net segmentation methods (2D, 3D) and SeqSeg using the Dice score ($\mathcal{D}$), Hausdorff distance ($\mathcal{H}$) and centerline overlap ($\mathcal{CO}$). The case types were either aortofemoral (AF) or aortic (AO), and the number of branches segmented is also shown (Nr. Br.). $^*$ indicates statistically significant difference ($p < 0.05$)}
\centering
\centerline{
\begin{tabular}{cccc|ccc|ccc|ccc}
    \toprule
    {} & {} & {} & {} & {} & {$\mathcal{D} \uparrow$} & {} & &{$\mathcal{H} \downarrow$} & {} & {} & {$\mathcal{CO} \uparrow$} & {} \\
    {Mod.} & {Case} & {Type} & {Nr.} & {} & {} & {} & {} & {} & {}& {} & {} & {}\\ 
    {} & {} & {} & {Br.} & {Seq-} & {2D} & {3D} & {Seq-} & {2D} & {3D} & {Seq-} & {2D} & {3D} \\ 
    {} & {} & {} & {} & {Seg} & {U-Net} & {U-Net} & {Seg} & {U-Net} & {U-Net}& {Seg} & {U-Net} & {U-Net}\\ 
    \midrule
  CT & 1 & AF & 9  & \textbf{0.907} & 0.885 & 0.846 &    \textbf{1.930} & 2.208 & 3.526 &    \textbf{0.939} & 0.906 & 0.657 \\
     & 2 & AF & 10 & 0.931 & \textbf{0.941} & 0.909 &    1.951 & \textbf{1.406} & 2.442 &    0.884 & \textbf{0.928} & 0.611 \\
     & 3 & AF & 10 & \textbf{0.885} & 0.860 & 0.855 &    \textbf{2.339} & 4.452 & 3.973 &    \textbf{0.959} & 0.864 & 0.791 \\
     & 4 & AO & 5  & 0.902 & \textbf{0.923} & 0.901 &    2.522 & \textbf{1.281} & 1.976 &    0.919 & \textbf{0.951} & 0.939 \\
     & 5 & AO & 5  & \textbf{0.940} & 0.845 & 0.865 &    \textbf{0.717} & 1.544 & 1.804 &    \textbf{1.000} & 0.916 & 0.992 \\
     & 6 & AO & 6  & \textbf{0.951} & 0.947 & 0.946 &    \textbf{0.867} & 0.991 & 0.999 &    \textbf{0.980} & 0.951 & 0.942 \\
     & 7 & AO & 5  & \textbf{0.955 }& 0.951 & 0.938 &    \textbf{0.725} & 3.237 & 3.216 &    \textbf{0.994} & 0.911 & 0.864 \\
     & 8 & AO & 4  & \textbf{0.954 }& 0.934 & 0.932 &    \textbf{0.708} & 1.358 & 1.473 &    0.990 & \textbf{1.000} & \textbf{1.000} \\
     & \textbf{Avg.} & - & - & \textbf{0.928} & 0.911 & 0.899 & \textbf{1.470} & 2.060 & 2.426 & \textbf{0.958} & 0.928 & 0.849 \\
     & {\footnotesize p-value} & - & - &  & {\footnotesize 0.547} & {\footnotesize 0.078} & & {\footnotesize 0.742} & {\footnotesize 0.148} & & {\footnotesize 0.547} & {\footnotesize 0.109} \\
     \midrule
  MR & 1 & AF & 9 & \textbf{0.877} & 0.706 & 0.816 &    \textbf{1.429} & 3.613 & 3.652 &    \textbf{0.977} & 0.510 & 0.650 \\
     & 2 & AO & 5 & \textbf{0.810} & 0.759 & 0.766 &    \textbf{1.154} & 1.515 & 2.337 &    \textbf{0.950} & 0.796 & 0.719 \\
     & 3 & AO & 5 & \textbf{0.836} & 0.824 & 0.822 &    \textbf{0.657} & 0.807 & 0.924 &    \textbf{0.844} & 0.816 & 0.711 \\
     & 4 & AO & 5 & 0.923 & \textbf{0.933} & 0.898 &    1.132 & \textbf{0.975} & 2.487 &    \textbf{0.981} & 0.929 & 0.855 \\
     & 5 & AO & 5 & \textbf{0.909} & 0.894 & 0.895 &    \textbf{1.371} & 2.141 & 2.048 &    \textbf{1.000} & 0.930 & 0.960 \\
     & 6 & AO & 5 & 0.923 & \textbf{0.932} & 0.920 &    0.704 & \textbf{0.653} & 0.986 &    \textbf{0.990} & 0.966 & 0.927 \\
     & 7 & AO & 4 & \textbf{0.945} & 0.925 & 0.904 &    \textbf{0.780} & 1.474 & 0.895 &    \textbf{1.000} & 0.929 & 0.983 \\
     & \textbf{Avg.} & - & - & \textbf{0.889} & 0.853 & 0.860 & \textbf{1.032} & 1.597 & 1.904 & \textbf{0.933} & 0.839 & 0.829 \\
     & {\footnotesize p-value} & - & - &  & {\footnotesize 0.078} & {\footnotesize 0.016$^*$} & & {\footnotesize 0.109} & {\footnotesize 0.016$^*$} & & {\footnotesize 0.016$^*$} & {\footnotesize 0.016$^*$} \\
    \hline
\end{tabular}
}
\label{tab:results_nokeep}
\end{table*}

\begin{table*}[t]
\caption{Quantitative comparison for the VMR test dataset \textit{after} largest connected body filtering between the two benchmark U-Net segmentation methods (2D, 3D) and our method, SeqSeg, using the Dice score ($\mathcal{D}$), Hausdorff distance ($\mathcal{H}$) and centerline overlap ($\mathcal{CO}$). The case types were either aortofemoral (AF) or aortic (AO), and the number of branches segmented is also shown (Nr. Br.). $^*$ indicates statistically significant difference ($p < 0.05$)}
\centering
\centerline{
\begin{tabular}{cccc|ccc|ccc|ccc}
    \toprule
    {} & {} & {} & {} & {} & {$\mathcal{D} \uparrow$} & {} & &{$\mathcal{H} \downarrow$} & {} & {} & {$\mathcal{CO} \uparrow$} & {} \\
    {Mod.} & {Case} & {Type} & {Nr.} & {} & {} & {} & {} & {} & {}& {} & {} & {}\\ 
    {} & {} & {} & {Br.} & {Seq-} & {2D} & {3D} & {Seq-} & {2D} & {3D} & {Seq-} & {2D} & {3D} \\ 
    {} & {} & {} & {} & {Seg} & {U-Net} & {U-Net} & {Seg} & {U-Net} & {U-Net}& {Seg} & {U-Net} & {U-Net}\\ 
    \midrule
    CT & 1 & AF &  9 & \textbf{0.907} & 0.879 & 0.830 &    \textbf{1.930} & 2.989 & 6.723 &    \textbf{0.939} & 0.824 & 0.521 \\
       & 2 & AF & 10 & 0.931 & \textbf{0.932} & 0.893 &    1.951 & \textbf{1.544} & 5.966 &    \textbf{0.884} & 0.800 & 0.439 \\
       & 3 & AF & 10 & \textbf{0.885} & 0.858 & 0.846 &    \textbf{2.339} & 4.810 & 5.060 &    \textbf{0.959} & 0.831 & 0.733 \\
       & 4 & AO &  5 & \textbf{0.902} & 0.887 & 0.916 &    \textbf{2.522} & 3.596 & 2.878 &    \textbf{0.919} & 0.869 & 0.865 \\
       & 5 & AO &  5 & \textbf{0.940} & 0.704 & 0.865 &    \textbf{0.717} & 8.163 & 1.804 &    \textbf{1.000} & 0.750 & 0.992 \\
       & 6 & AO &  6 & \textbf{0.951} & 0.945 & 0.946 &    \textbf{0.867} & 1.335 & 0.999 &    \textbf{0.980} & 0.865 & 0.942 \\
       & 7 & AO &  5 & 0.955 & \textbf{0.952} & 0.939 &    \textbf{0.725} & 2.464 & 2.875 &    \textbf{0.994} & 0.907 & 0.831 \\
       & 8 & AO &  4 & \textbf{0.954} & 0.934 & 0.932 &    \textbf{0.708} & 1.358 & 1.473 &    0.990 & \textbf{1.000} & \textbf{1.000} \\
     & \textbf{Avg.} & - & - & \textbf{0.928} & 0.886 & 0.896 & \textbf{1.470} & 3.282 & 3.472 & \textbf{0.958} & 0.856 & 0.790 \\
     & {\footnotesize p-value} & - & - &  & {\footnotesize 0.109} & {\footnotesize 0.109} & & {\footnotesize 0.109} & {\footnotesize 0.039$^*$} & & {\footnotesize 0.023$^*$} & {\footnotesize 0.039$^*$} \\
    \midrule
    MR & 1 & AF & 9 & \textbf{0.877} & 0.508 & 0.810 &    \textbf{1.429} & 13.575 & 3.652 &    \textbf{0.977} & 0.199 & 0.623 \\
       & 2 & AO & 5 & \textbf{0.810} & 0.752 & 0.766 &    \textbf{1.154} & 1.874 & 2.337 &    \textbf{0.950} & 0.744 & 0.713 \\
       & 3 & AO & 5 & \textbf{0.836} & 0.832 & 0.822 &    \textbf{0.657} & 0.807 & 0.924 &    \textbf{0.844} & 0.816 & 0.711 \\
       & 4 & AO & 5 & 0.923 & \textbf{0.931} & 0.898 &    \textbf{1.132} & 1.743 & 2.524 &    \textbf{0.981} & 0.894 & 0.855 \\
       & 5 & AO & 5 & \textbf{0.909} & 0.897 & 0.898 &    \textbf{1.371} & 1.925 & 1.778 &    \textbf{1.000} & 0.930 & 0.960 \\
       & 6 & AO & 5 & 0.923 & \textbf{0.930} & 0.920 &    \textbf{0.704} & 0.831 & 0.986 &    \textbf{0.990} & 0.916 & 0.927 \\
       & 7 & AO & 4 & \textbf{0.945} & 0.928 & 0.904 &    \textbf{0.780} & 1.540 & 0.895 &    \textbf{1.000} & 0.929 & 0.983 \\
     & \textbf{Avg.} & - & - & \textbf{0.889} & 0.825 & 0.860 & \textbf{1.032} & 3.185 & 1.871 & \textbf{0.963} & 0.776 & 0.825\\
     & {\footnotesize p-value} & - & - &  & {\footnotesize 0.156} & {\footnotesize 0.016$^*$} & & {\footnotesize 0.016$^*$} & {\footnotesize 0.016$^*$} & & {\footnotesize 0.016$^*$} & {\footnotesize 0.016$^*$} \\
    
    \hline
\end{tabular}
}
\label{tab:results_keep}
\end{table*}

\begin{table*}[t]
\caption{Quantitative comparison for the AVT dataset between the benchmark 2D U-Net segmentation method, raw output and after largest connected body filtering (LC), and SeqSeg using the Dice score ($\mathcal{D}$), Hausdorff distance ($\mathcal{H}$) and centerline overlap ($\mathcal{CO}$). The number of branches segmented is shown (Nr. Br.) and $^*$ indicates statistically significant difference ($p < 0.05$)}
\centering
\centerline{
\begin{tabular}{ccc|ccc|ccc|ccc}
    \toprule
    {} & {} & {} & {} & {$\mathcal{D} \uparrow$} & {} & &{$\mathcal{H} \downarrow$} & {} & {} & {$\mathcal{CO} \uparrow$} & {} \\
    {Mod.} & {Case} & {Nr.} & {} & {} & {} & {} & {} & {}& {} & {} & {}\\ 
    {} & {} & {Br.} & {Seq-} & {2D} & {LC 2D} & {Seq-} & {2D} & {LC 2D} & {Seq-} & {2D} & {LC 2D} \\ 
    {} & {} & {} & {Seg} & {U-Net} & {U-Net} & {Seg} & {U-Net} & {U-Net}& {Seg} & {U-Net} & {U-Net}\\ 
    \midrule
CT & 1 & 15 & \textbf{0.924} & 0.87 & 0.843 &    67.9 & \textbf{51.7} & 172 &    \textbf{0.782} & 0.506 & 0.301 \\
& 2 & 14 & \textbf{0.951} & 0.901 & 0.881 &    \textbf{28.2} & 42.6 & 36.9 &    \textbf{1} & 0.899 & 0.813  \\
& 3 & 14 & \textbf{0.925} & 0.567 & 0.299 &    60.4 & \textbf{56.8} & 132 &    \textbf{0.929} & 0.568 & 0.338 \\
& 4 & 10 & \textbf{0.912} & 0.857 & 0.858 &    \textbf{17.4} & 29.9 & 29.9 &    \textbf{1} & 0.967 & 0.966 \\
& 5 & 10 & \textbf{0.898} & 0.823 & 0.758 &    \textbf{27.2} & 30.4 & 96.5 &    \textbf{0.976} & 0.834 & 0.581  \\
& 6 & 8 & \textbf{0.87} & 0.854 & 0.835 &    38.8 & \textbf{29.7} & 44.7 &    \textbf{0.973} & 0.928 & 0.783  \\
& 7 & 13 & \textbf{0.936} & 0.787 & 0.778 &    \textbf{26.4} & 39.2 & 42.7 &    \textbf{0.933} & 0.887 & 0.812  \\
& 8 & 8 & \textbf{0.911} & 0.857 & 0.799 &    25 & \textbf{17.7} & 130 &    \textbf{0.981} & 0.743 & 0.412  \\
& 9 & 8 & 0.875 & 0.884 & \textbf{0.885} &    26.3 & \textbf{23.3} & 22.3 &    0.988 & \textbf{0.989} & \textbf{0.989}  \\
& 10 & 11 & 0.872 & \textbf{0.874} & 0.868 &    99.6 & \textbf{52.4} & 96.4 &    0.757 & \textbf{0.874} & 0.764  \\
& 11 & 11 & \textbf{0.935} & 0.86 & 0.795 &    19.9 & \textbf{19.7} & 120 &    \textbf{0.98} & 0.705 & 0.459 \\
& 12 & 8 & \textbf{0.892} & 0.846 & 0.846 &    \textbf{22} & 29.4 & 29.4 &    \textbf{0.992} & 0.969 & 0.969  \\
& 13 & 9 & \textbf{0.953} & 0.892 & 0.867 &    \textbf{12.8} & 23.9 & 72.2 &    \textbf{0.93} & 0.815 & 0.502  \\
& 14 & 5 & \textbf{0.918} & 0.883 & 0.886 &    \textbf{23.4} & 26.8 & 26.8 &    \textbf{1} & 0.992 & 0.992  \\
& 15 & 7 & \textbf{0.916} & 0.902 & 0.899 &    30.1 & 22.5 & \textbf{18.8} &    \textbf{0.995} & 0.978 & 0.894  \\
& 16 & 8 & \textbf{0.914} & 0.866 & 0.858 &    \textbf{17.3} & 26.4 & 29.3 &    \textbf{0.995} & 0.966 & 0.966  \\
& 17 & 11 & \textbf{0.904} & 0.863 & 0.792 &    \textbf{18.6} & 25.6 & 69.2 &    \textbf{0.983} & 0.796 & 0.572  \\
& 18 & 6 & \textbf{0.941} & 0.897 & 0.867 &    \textbf{12.9} & 18.8 & 75.5 &    \textbf{0.998} & 0.837 & 0.632  \\
& \textbf{Avg.} & - & \textbf{0.914} & 0.849 & 0.812 & 31.899 & \textbf{31.498} & 69.120 & \textbf{0.955} & 0.847 & 0.708\\
     & {\footnotesize p-value} & - &  & {\footnotesize 5.3e-4$^*$} & {\footnotesize 3.8e-5$^*$} & & {\footnotesize 0.609} & {\footnotesize 4.2e-4$^*$} & & {\footnotesize 3.8e-5$^*$} & {\footnotesize 2.3e-5$^*$} \\
    
    \hline
\end{tabular}
}
\label{tab:results_avt}
\end{table*}

\section{Discussion}

U-Net learning models, and particularly the more recent nnU-Net, have shown excellent potential for automating image segmentation tasks. However, segmentation of branched vascular structures from medical image data is fraught with challenges since vessels typically compose relatively few pixels, vascular geometry varies considerably between patient and location, and maintaining connectivity of highly branched structures by pixel classification is tricky. We herein propose a sequential segmentation technique (SeqSeg) that leverages U-Net learning to locally build vascular models. We observed that SeqSeg generally outperformed current state-of-the-art global nnU-Net models when tested on typical vascular images used for image-based modeling, particularly when comparing overall extent of connected vasculature predicted. Namely, SeqSeg was superior in extending the segmentations into smaller branch arteries or distal segments when compared to the global nnU-Net benchmarks (or, in fact, ``ground truth'' segmentations). In addition, SeqSeg performance was generally more robust, as indicated by less variance in the quantitative results. 

We note that when comparing the difference in the quantitative metrics across the VMR test cases, the superiority of SeqSeg did not necessarily reach statistical significance as measured by $p<0.05$. This is perhaps due to the smaller number of test cases we had access to. In addition, the VMR ground truth used for evaluating these metrics were not as extensive as they could have been, which likely handicapped the comparison since SeqSeg generally excelled at extending the segmentations further down the vascular tree compared to the nnU-Net benchmarks. Namely, the ground truth data tended to be limited to the aorta and proximal segments of branch arteries, which are generally easier to segment, and both SeqSeq and the nnU-Net benchmarks performed {\em on average} equally well in such ``less-challenging'' regions. Moreover, metrics like Dice are inherently biased to larger vessels. While the ground truth segmentations could have been manually altered to extend vessels, or include missing branches, this can introduce potential bias; thus, we chose not to adulterate the ground truth from the public repository. Lastly, we note that most prior publications in this field \citep{NvidiaUNETR:Segmentation,ChenTransUNet:Segmentation,DosovitskiyANSCALE,Milletari2016V-Net:Segmentation, Chen2019Med3D:Analysis, Hatamizadeh2022UNetFormer:Segmentation} do not report whether their improvements to prior benchmarks were statistically significant. 

The application of SegSeg to the AVT CT test dataset offered an interesting application. Whereas SegSeg and the benchmarks were trained on VRM data, with a subset of the VMR data held out for testing, the AVT data was a completely {\em de novo} data source unrelated to the training. For the application to this {\em de novo} data, SegSeg more convincing outperformed the nnU-Net benchmarks, achieving statistically significant higher accuracy in terms of {\em all} quantitative metrics. Moreover, the qualitative comparisons shown in Fig.~\ref{fig:qual_avt} demonstrate that SegSeq was able to segment far more aortic branch arteries, and extend arteries further distally than the benchmark, or, in fact, even than the ground truth. It is unknown if this superior performance extends to similar {\em de novo} MR data since such data was not available.        

The advantage of SeqSeg is that it focuses the segmentation task locally around a vascular segment. Indeed, the same neural network architecture and training strategies were used for SeqSeg and the benchmarks. The centerline overlap metric, which give increased weight to smaller branches and bifurcations, showed generally better performance for SeqSeg. In fact, after largest connected region filtering, the difference in centerline overlap metric was statistically significant for all test datasets. This is likely because the nnU-Net benchmarks may sacrifice smaller branches in order to accurately predict larger branches that carry more pixels. In contrast, SeqSeg deals with all branches, if detected, more equitably, by focusing on one segment at a time. Figures~\ref{fig:qual_vmr_ct},~\ref{fig:qual_vmr_mr}, and ~\ref{fig:qual_avt} show another major advantage of assembling a vascular network piece-wise: ensuring connectivity of the resulting model, which is crucial for blood flow and tissue mechanics simulation purposes.

Since the overall objective is to produce models capable of physics based simulation, it is notable to mention SeqSegs superiority towards that goal. Firstly, as mentioned above, SeqSeg surpasses the benchmark's' ability to generate expansive {\em and} single connected body models. Secondly, since SeqSeg traces the vasculature, it maintains information on branches and their connections relative to the global vascular organization. This can be used to place boundary conditions (inlet and outlet conditions), necessary for physics-based simulation setup. This information is not available for typical CNN segmentation methods since all pixels are treated equally and vasculature organization is ignored. The authors note that this study does not directly look at the effects of these methods on actual physics-based simulation results, which is beyond the scope of the current study.

One might assume an advantage of SegSeq is that because it uses local patches of the image volume, the number of inputs for training is higher than for the global nnUnets. However, during training nnUnet partitions the image volume into patches and uses extensive augmentations, which greatly increases the ``samples'' for training.  

The results from the benchmark 2D and 3D nnU-Nets show the limitations of 3D convolutional neural networks for global vasculature segmentation--the problems of class-imbalance and image size. Because global image volumes surpass GPU memory, methods are forced to either downsample or split the image into patches to fit on a GPU. Our method excels within the constraints of GPU memory because it processes smaller subvolumes at each time, which rarely exceeds the GPU memory limit, see Table~\ref{tab:nnunet} for larger possible batch sizes for example. Furthermore, in a global image volume, the vascular pixels represent only a fraction of the total pixels, making training difficult. Our method focuses on the pixels around the vasculature, which, by definition, alleviates class imbalance.

On the other hand, the results also indicate that the source of better segmentation is not simply locality. The benchmark models were trained on small patches that do not undergo downsampling. By training locally {\em and} incorporating prior learned information, i.e. the location and size of the vessel, SeqSeg is generally able to segment with greater detail and accuracy, particularly in smaller vessels. 

Another limitation that impacts global segmentation learning is that ground truth segmentation, being human-generated, in most cases did not contain segmentation of all branches or portions. This implies that some training data had certain arteries, e.g., the renal arteries, segmented while others did not, which can result in poor segmentation of test data. Since SeqSeg can utilize training patches around vessels, the training mostly encounters positive examples of arteries, e.g. the renal arteries only if they are present, and will not encounter negative (wrong) examples from less segmented images, e.g. where the renal arteries were not segmented. Thus, SeqSeg can be more efficient with training data, which is beneficial since annotated data collection is costly and time-consuming. This could also help explain the ability of SeqSeg to segment a greater number of smaller branches, even those not present in all training examples.

Additionally, SeqSeg may have been able to generalize to regions not present in the training data because vessels share similar image features when viewed locally. For example, Figures~\ref{fig:qual_vmr_ct} and~\ref{fig:qual_vmr_mr} show how SeqSeg managed to trace and segment small bifurcations not included in the ground truth as well as elongate already segmented vessels. Inspection confirmed that these vessels were present in the image data. In fact, the authors further confirmed this qualitatively by training a model solely on {\em one} branch per case (the aorta), and SeqSeg was able to generalize to other branches on test data. Furthermore, SeqSeg manages to detect and handle bifurcations, which has been an challenge for blood vessel tracking and tracing methods \citep{Jia2021Learning-basedReview, Wolterink2019CoronaryClassifier,  Abbasi-Sureshjani2016AutomaticScores, Pratt2018AutomaticPhotography, Li2022VBNet:Images}. Unlike other works, SeqSeg does not depend on explicit bifurcation detection, but instead handles them implicitly through surface representations expressing them. This makes handling complex junctions with multiple branches more tractable.

For further comparison to previous works, SegSeg achieved better global Dice scores than Maher et al.\ who trained neural networks for 2D lumen segmentation on similar datasets \citep{Maher2020NeuralModeling}. Furthermore, the method of Maher et al.\ depended on previously user-generated centerlines, whereas our method automatically generates the centerlines while simultaneously segmenting the vasculature. This is significant since centerline generation is often the most time-consuming step of image-based model construction.

SegSeg used a 3D U-Net neural network architecture for local segmentation predictions, however, other architectures, such as vision transformers \citep{Hatamizadeh2022UNetFormer:Segmentation}, transfer learning models such as 3D MedNet \citep{Chen2019Med3D:Analysis} or V-Net \cite{Milletari2016V-Net:Segmentation}, with residual connections, could possibly be used to perform this task. Similarly, future developments could include learning methods to determine step size or other parameters that are derived from deterministic procedures in our presented model. For example, deep learning can be applied to directly predict subsequent points~\citep{Dorobantiu2021Coronary3d-unet}, local centerline segments or surfaces using template-based approaches similar to what has been done for cardiac model construction~\citep{Kong2021AReconstruction}. Additionally, the SeqSeg method can be trained and tested for generalization to other vascular anatomies such as coronary arteries, pulmonary arteries and cerebrovascular models. Since the training and testing occur locally, new data from different anatomies can be readily incorporated into the existing framework.

Limitations of the presented method include the dependence on voxel-based segmentation, the dependence on capturing bifurcation roots, and the possible computational cost. Voxel-based segmentation inevitably leads to staircase artifacts on the final surface, as shown in Figures~\ref{fig:qual_vmr_ct} and \ref{fig:qual_vmr_mr}. Since our stepwise approach relies on accurately capturing bifurcation roots, there is the possibility of missing whole branches if the root is difficult to segment, e.g. because of image artifacts. The method also requires neural network inference at each step, which has the potential to scale poorly for extensive vascular networks.

\section{Conclusion}
Despite its growing importance, reconstructing vascular models from medical image data in an accurate and rapid manner remains an open area of research. In this work we present SeqSeg; a novel image-based vascular model construction method based on building the vascular network in a stepwise manner to facilitate learning. SeqSeg is capable of automatically tracing and assembling a global segmentation and surface, depending only on a single seed point. We tested the method on CT and MR images of aortic and aortofemoral models and compared to state-of-the-art benchmark 2D and 3D U-Net segmentation methods, SegSeg had similar or better accuracy in terms of Dice score, Hausdorff distance, and centerline overlap, but more notably was more robust and able to connect a greater extent of the vasculature. Our future work includes training and testing using other vascular anatomies as well as further optimizing local segmentation and bifurcation detection.

\bmhead{Acknowledgments}

We thank Prof. Alison Marsden and Dr. Nathan Wilson for their input to this work, Dr.\ Fanwei Kong for advice, comments and support early on and we thank Dr.\ Martin Pfaller for valuable input and dataset curation and access. This research used the Savio computational cluster provided by the Berkeley Research Computing program at the University of California, Berkeley.

\section*{Declarations}
\paragraph{Funding} Earlier stages of this work were supported by the National Institute of Health, Award No.~5R01LM013120. We also acknowledge support by the National Science Foundation, Award No.~1663747 and 2310910 for later stages of this work.
\paragraph{Conflict of interest} The authors have no competing interests to declare that are relevant to the content of this article.
\paragraph{Availability of data and materials} Data used for this study is available at \url{vascularmodel.com}
\paragraph{Code availability} Code written and used for this manuscript is available at \url{https://github.com/numisveinsson/SeqSeg}
\paragraph{Authors' contributions} NSC and SCS conceptualized the study design and methods. NSC developed the algorithms, performed the computations and generated results. NSC and SCS interpreted the results. NSC developed the first draft of the manuscript and SCS further contributed to and revised the manuscript. 

\paragraph{Citation Diversity Statement}
Recent work in several fields of science has identified a bias in citation practices such that papers from women and other minority scholars are undercited relative to the number of papers in the field \citep{Caplar2016QUANTITATIVECOUNTS, Maliniak2013TheRelations, Mitchell2013GenderedJournals}. We recognize this bias and have worked diligently to ensure that we are referencing appropriate papers with fair gender and racial author inclusion.

\bibliography{references}

\begin{appendices}

\section{}

\begin{figure}[h]
    \centering
    \subfigure[]{\includegraphics[width=0.48\textwidth]{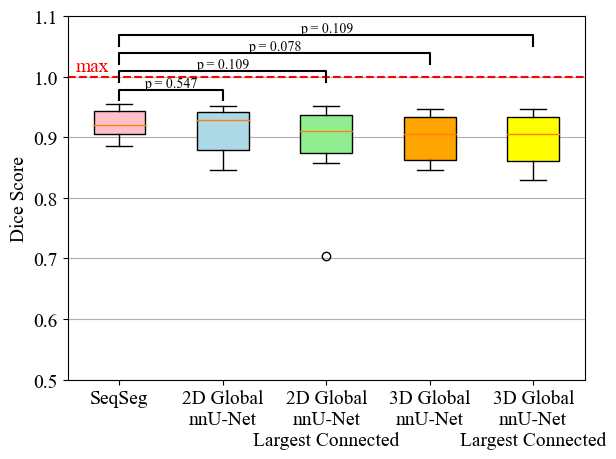}} 
    \subfigure[]{\includegraphics[width=0.48\textwidth]{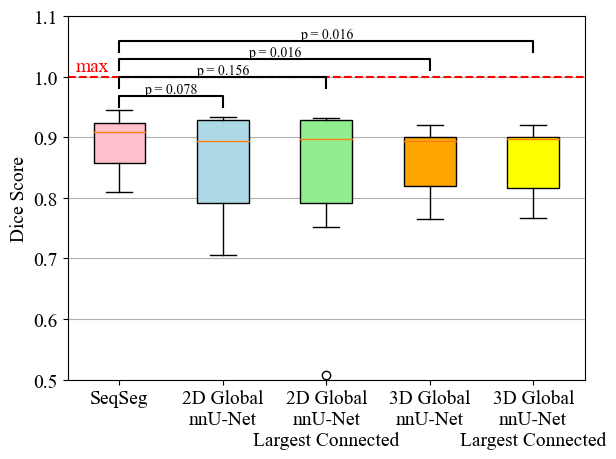}} 
    \subfigure[]{\includegraphics[width=0.48\textwidth]{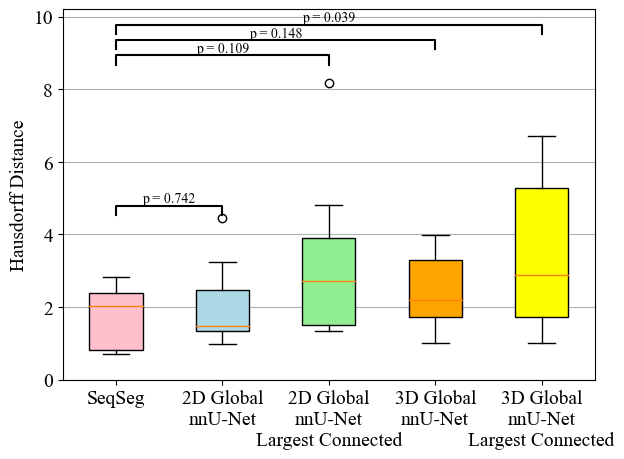}}
    \subfigure[]{\includegraphics[width=0.48\textwidth]{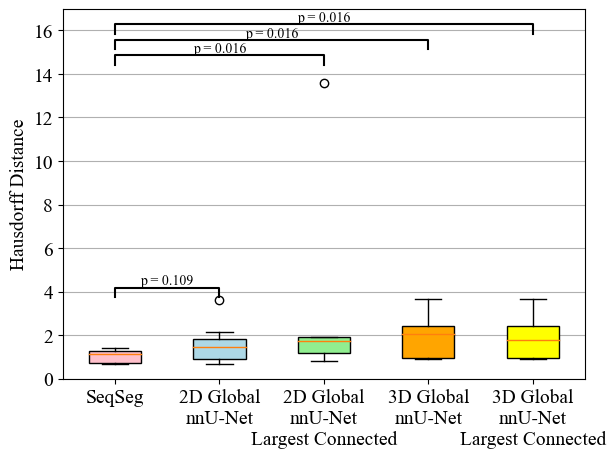}}
    \subfigure[]{\includegraphics[width=0.48\textwidth]{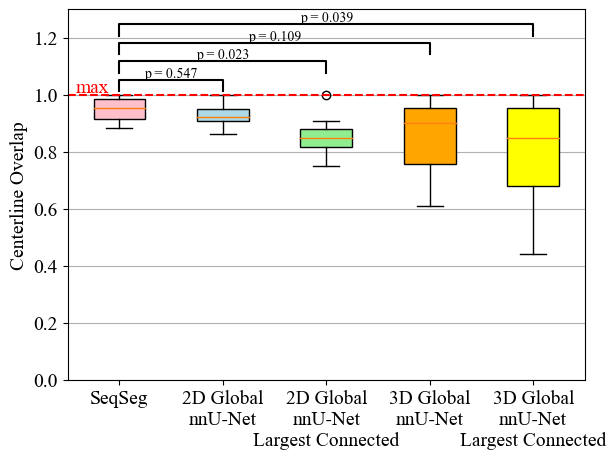}}
    \subfigure[]{\includegraphics[width=0.48\textwidth]{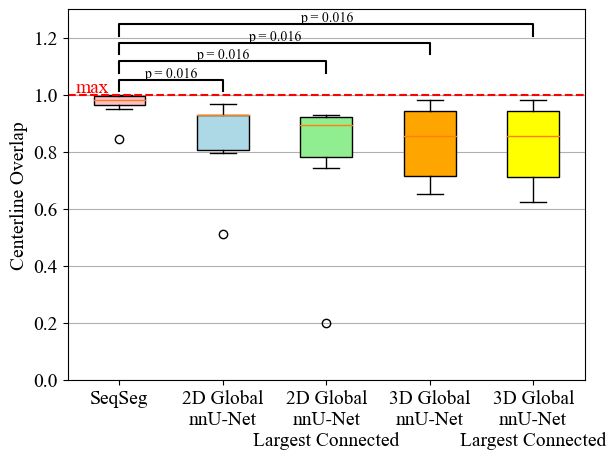}}
    \caption{ Quantitative metrics for VMR test data; (a) CT Dice (b) MR Dice (c) CT Hausdorff (in pixels) (d) MR Hausdorff (in pixels) (e) CT Centerline Overlap (f) MR Centerline Overlap. See Sec.~\ref{sec:exp} for definition of p-values.  
    }
    \label{fig:results_vmr_all}
\end{figure}

\begin{figure}[h]
\begin{center}
\includegraphics[width=0.8\linewidth]{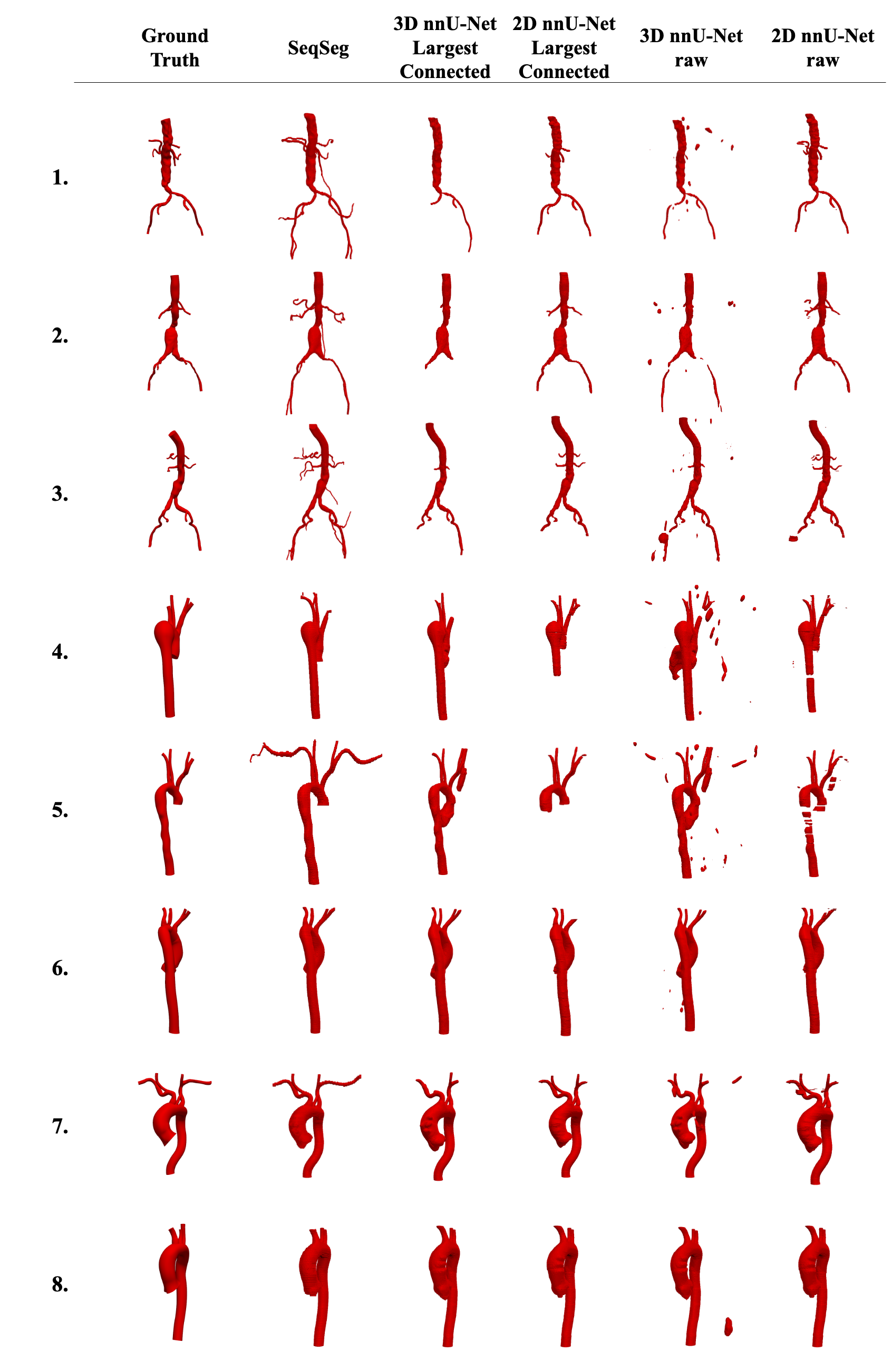}
\end{center}
\caption{ Resulting meshes from complete VMR CT test dataset. Each row represents a different vascular model, labelled consistently with Tables~\ref{tab:results_nokeep},~\ref{tab:results_keep}}\label{fig:qual_vmr_ct}
\end{figure}

\begin{figure}[h]
\begin{center}
\includegraphics[width=0.8\linewidth]{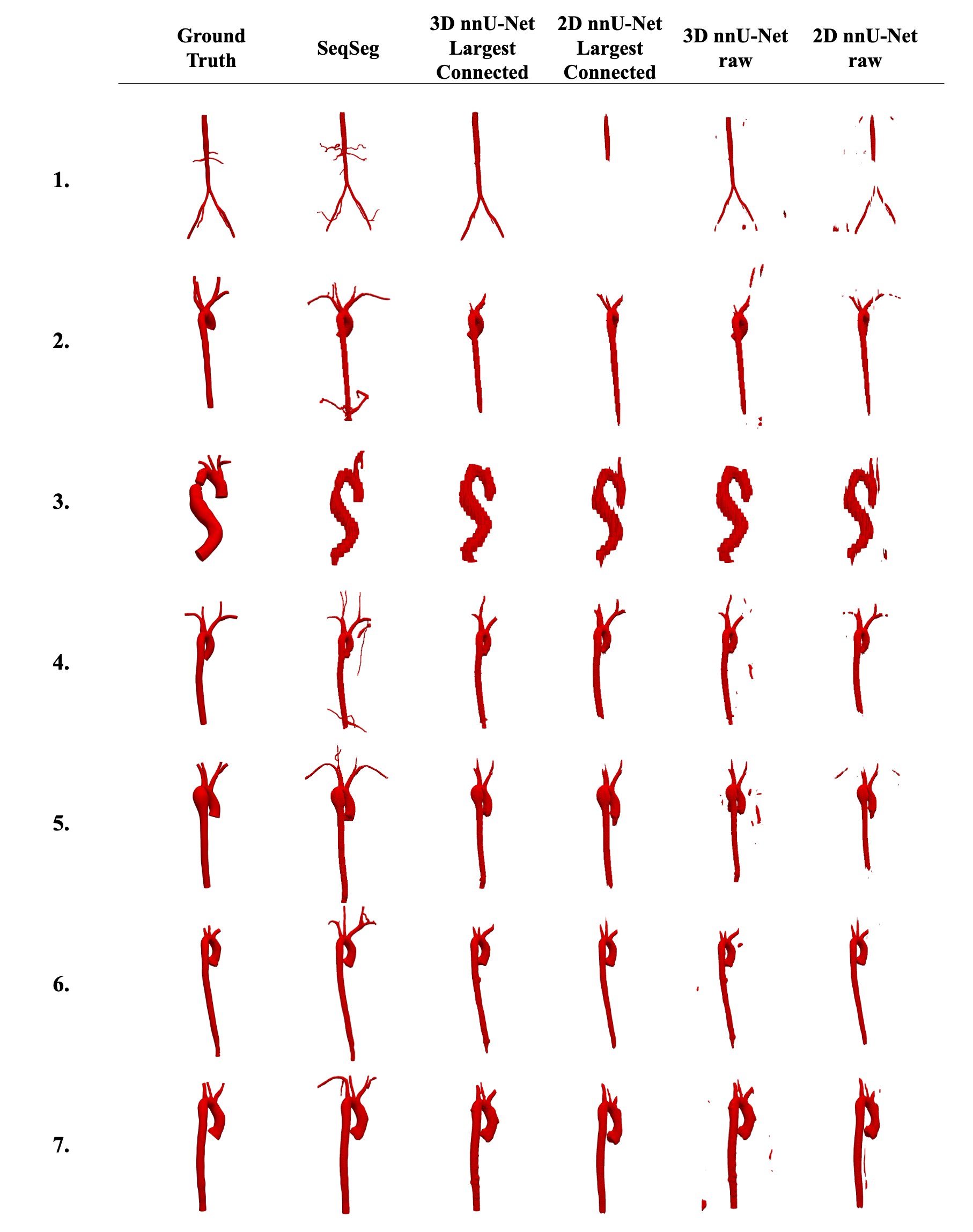}
\end{center}
\caption{ Resulting meshes from complete VMR MR test dataset. Each row represents a different vascular model, labelled consistently with Tables~\ref{tab:results_nokeep},~\ref{tab:results_keep}}\label{fig:qual_vmr_mr}
\end{figure}

\begin{figure}[h]
\begin{center}
\includegraphics[width=0.99\linewidth]{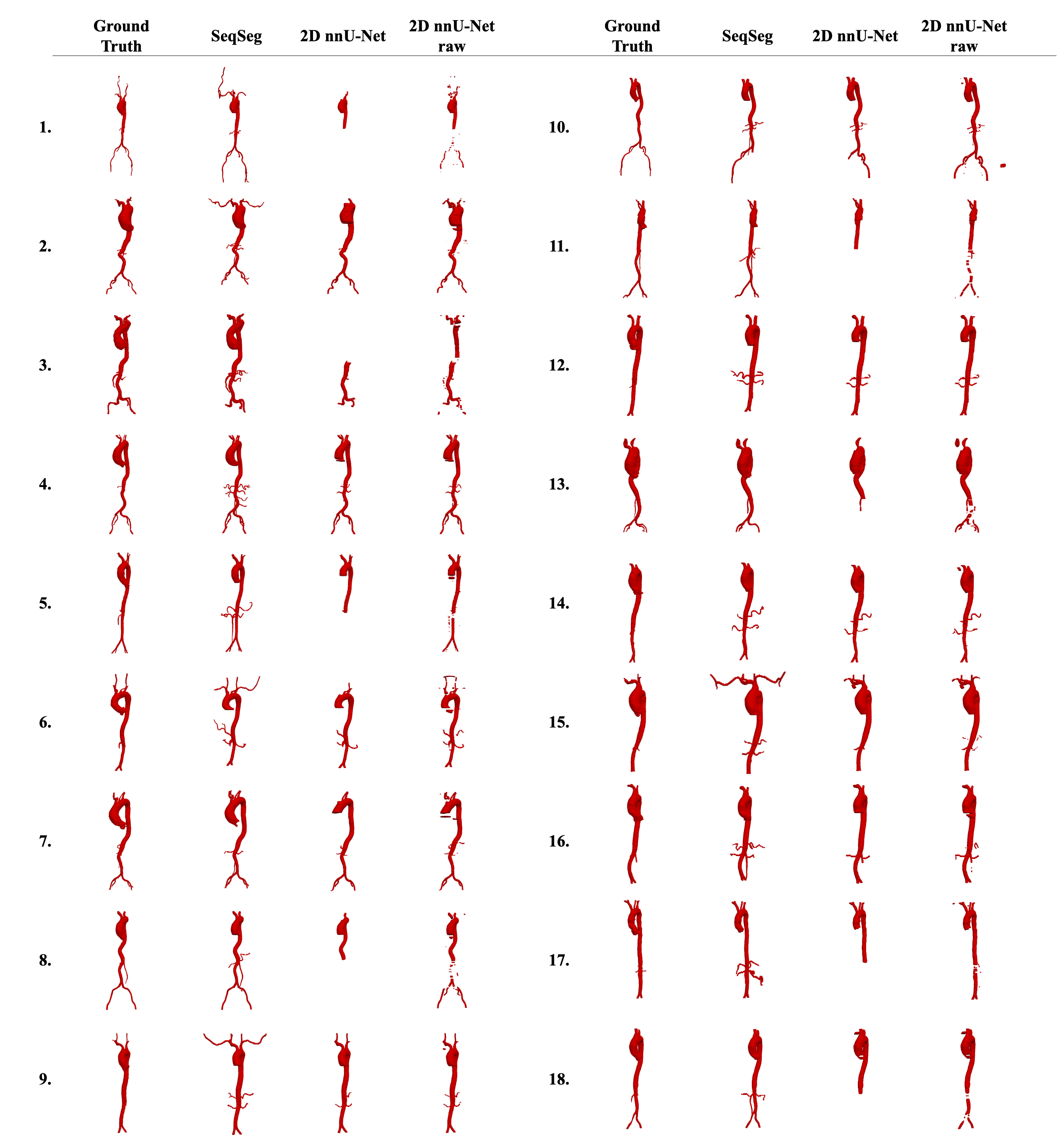}
\end{center}
\caption{ Resulting meshes from complete AVT CT test dataset. Each row represents a different vascular model, labelled consistently with Table~\ref{tab:results_avt}}\label{fig:qual_avt}
\end{figure}

\end{appendices}

\end{document}